\documentclass[12pt, a4paper]{article}

\usepackage{amsfonts}
\usepackage{amsmath}
\usepackage{amssymb}
\usepackage{amsthm}
\usepackage{mathtools}
\usepackage{authblk}
\usepackage{appendix}
\usepackage{bm}
\usepackage{bbm}
\usepackage{dsfont}
\usepackage{natbib}
\usepackage{subcaption}
\usepackage{lineno}
\usepackage[hidelinks]{hyperref}
\usepackage{enumitem}
\usepackage{booktabs}
\usepackage[T1]{fontenc}
\usepackage{multirow}
\usepackage{soul}
\usepackage{xcolor}
\usepackage[font = small]{caption}

\usepackage{graphicx,url}

\newtheorem{theorem}{Theorem}
\newtheorem{corollary}{Corollary}
\newtheorem{lemma}{Lemma}

\newtheorem{assumption}{Assumption}
\newtheorem{proposition}{Proposition}

\theoremstyle{definition}

\usepackage[utf8]{inputenc}

\newcommand{\norm}[1]{\left\lVert#1\right\rVert}

\graphicspath{ {../images/} }

\newcommand{\blind}{1}
\newcommand{\spacing}{1.1}

\begin{document}

\def\spacingset#1{\renewcommand{\baselinestretch}%
{#1}\small\normalsize} \spacingset{1}

\if1\blind
{
  \title{\bf Posterior Projection for Inference in Constrained Spaces}
    \author[1,2]{Lachlan Astfalck}
    \author[3]{Deborshee Sen}
    \author[3]{Sayan Patra}
    \author[2]{Edward Cripps}
    \author[3]{David Dunson}
    \affil[1]{School of Mathematics \& Statistics, University of New South Wales, Australia}
    \affil[2]{School of Physics, Mathematics \& Computing, University of Western Australia, Australia}
    \affil[3]{Department of Statistical Science, Duke University, Durham, NC, USA}
    \setcounter{Maxaffil}{0}
    \renewcommand\Affilfont{\itshape\small}
  \maketitle
} \fi

\if0\blind
{
  \bigskip
  \bigskip
  \bigskip
  \begin{center}
    {\LARGE\bf Posterior Projection for Inference in Constrained Spaces}
\end{center}
  \medskip
} \fi

\bigskip
\begin{abstract}
  Estimation of parameters that obey specific constraints is crucial in statistics and machine learning; for example, when parameters are required to satisfy boundedness, monotonicity, or linear inequalities. Traditional approaches impose these constraints via constraint-specific transformations or sampling approaches, or by truncating the posterior distribution. Such methods often result in computational challenges, limited flexibility, and a lack of generality. We propose a generalized framework for constrained Bayesian inference by projecting the unconstrained posterior distribution into the space of the parameter constraints, providing a computationally efficient and easily implementable solution for a large class of problems. We rigorously establish the theoretical foundations of the projected posterior distribution, as well as providing asymptotic results for posterior consistency, posterior contraction, and optimal coverage properties. Our methodology is validated through both theoretical arguments and practical applications.
\end{abstract}

\noindent%

\spacingset{\spacing}

\section{Introduction} \label{sec:intro}

Since the early publications of the Monte Carlo method \citep{Metropolis01091949,von195113} to the end of the 20th century, generating
pseudo-random variates has been central to the development of probability, statistics and machine learning. Non-uniform sampling methods accelerated Monte Carlo sampling \citep{devroye1986nonuniform}, in turn facilitating the implementation and development of Markov chain Monte Carlo (MCMC) methods 
proposed in the seminal articles of \cite{metropolis1953equation}, \cite{hastings1970monte}, \cite{geman1984stochastic}, 
\cite{tanner1987calculation} and \cite{gelfand1991gibbs}. The realization, brought upon by MCMC, of the possibility to sample from an extremely large class of (non-standard) distributions elevated Bayesian inference to the position it holds in statistics and machine learning today \citep[for a summary, see][]{brooks2011MCMC}. Nevertheless, despite advances in computer hardware, software and optimization tools experienced in the 21st century, the increasing complexity of modern real-world problems necessitates complex models that in practice remain difficult to sample from and so efficient solutions and implementations are required. More recently, much work has been dedicated to addressing the challenges with simulating from increasingly intricate constrained random variables. In what follows, we describe a unifying theory for parameter estimation in constrained spaces via a projection of an {\em unconstrained posterior} into the constrained space. It is a method that can be applied to a wide class of constraints, has theoretical foundations and, importantly, is computationally expedient and is easy to implement.

Strategies for enforcing parameter constraints typically fall into three broad paradigms. First, reparameterizations map a constrained space to an unconstrained domain. Canonical examples include log and log-link transformations for positive parameters in generalized linear models, logit or softmax mappings for probabilities and simplex-constrained weights \citep[e.g.][]{aitchison1982statistical,mccullagh1989generalized}. Second, constraints may be handled directly at the level of the sampler via constrained Monte Carlo methods that preserve the constrained geometry without reparameterizing the space. For example, truncated or reflective Hamiltonian Monte Carlo \citep[e.g.][]{pakman2014exact,kook2022sampling}, Gibbs sampling on the admissible space \citep[e.g.][]{ishwaran2001gibbs,gelman2013bayesian}, linear-inequality-constrained Gaussian models \citep[e.g.][]{jidling2017linearly,harkonen2023gaussian,dalton2024}, and Langevin-based algorithms \citep[e.g.][]{bubeck2018sampling,ahn2021efficient}. Notably, recent work on proximal mappings have been proposed as a unifying framework for constrained and regularized inference by incorporating constraints and non-smooth penalties through proximal operators \citep[see][]{pereyra2016proximal,xu2024bayesian,zhou2024proximal}. While conceptually appealing, proximal mapping methods still require explicit implementation of problem-specific samplers. Third, when the parameters live in a space with a lower topological dimension than the ambient space, such as the probability simplex or directional and spherical domains, one may instead formulate inference directly on the corresponding manifold, employing geometry-aware algorithms such as Riemannian or geodesic Monte Carlo \citep{girolami2011riemann,byrne2013geodesic}. Each paradigm introduces methodology tailored to the specific structural features of the constraint set, rather than a unified mechanism for incorporating constraints at the level of the posterior distribution itself. 

To formalize, denote by $\theta$ an unknown random variable of dimension $d$ with prior probability measure $\Pi_\Theta(\theta)$ over the (unconstrained) support $\Theta \in \mathbb{R}^d$. A general methodology for specifying a prior measure on a constrained space $\tilde{\Theta} \subseteq \Theta$ is to let $\Pi_{\tilde{\Theta}}(\theta) \propto \Pi_{\Theta}(\theta) \mathds{1}\{\theta \in \tilde{\Theta}\}$ where $\mathds{1}\{\cdot\}$ is the indicator function. Given an observed set of data $x_{(n)} = (x_1, \dots, x_n)$, and likelihood $\mathrm{P}(x_{(n)}|\theta)$ such a prior specification leads to the posterior distribution
\begin{equation} \label{eqn:truncated}
  \begin{split}
    \Pi_{\tilde{\Theta}}(\theta \mid x_{(n)}) & \propto \mathrm{P}(x_{(n)}|\theta)\Pi_\Theta(\theta)\mathds{1}\{\theta \in \tilde{\Theta}\} \\
  & \propto \Pi_{\Theta}(\theta \mid x_{(n)}) \mathds{1}\{\theta \in \tilde{\Theta}\}.
  \end{split}
\end{equation}
We name constrained posterior distributions on the left-hand side of \eqref{eqn:truncated} \textit{truncated} posterior distributions and note that they are rarely analytically available and typically numerically approximated via rejection sampling from the unconstrained posterior distribution \citep{rao2016data}. Although in theory this offers a general solution to the problem of parameter constraints, rejection sampling becomes difficult to implement when $\Pi_{\Theta}(\theta \mid x_{(n)})$ has much of its probability mass outside of $\tilde{\Theta}$, a challenge that is exacerbated with increasing dimensionality $d$. When $\tilde{\Theta}$ is of lower topological dimension than $\Theta$, and so has measure-zero, rejection sampling does not work altogether: for example, when $\theta$ denotes a compass bearing and so may be modeled as one-dimensional hyper-sphere $\tilde{\Theta} = \mathbb{S}^1 \subset \mathbb{R}^2$ in the real-valued plane $\Theta = \mathbb{R}^2$. In contrast, we propose posterior projection, a methodology that enforces parameter constraints by projecting samples from $\Pi_\Theta(\theta \mid x_{(n)})$ into $\tilde{\Theta}$.

The foundations of posterior projections are not new, and recent work has shown its potential.  \cite{dunson2003bayesian} and \cite{gunn2005transformation} consider order constrained parameters in generalized linear models and \cite{silva2015bayesian} study constraints for covariance matrices of latent Gaussian models for probit regression. The method is extremely well suited to constraints on GPs, as shown in \cite{lin2014bayesian} and \cite{chakraborty2021convergence} who present posterior projection methods to restrict GPs to be monotonic and \cite{wang2014modeling} who restrict GPs to directional quantities, similar to the compass bearing example above. In these works, the projected posteriors all have better empirical and finite sample performance than the comparative truncated posteriors. Despite the appeal of the projected posterior approach, it has only been implemented for specific cases, as mentioned above, and it lacks justification as a general theory.

We rectify these gaps by presenting a rigorous and general framework for Bayesian inference via posterior projection. Unlike previous approaches, our methodology and theory are not tailored to a specific constraint but instead apply broadly across a wide class of parameter restrictions, including monotonicity, boundedness, positivity, and manifold-valued constraints.  Furthermore, we permit projection under any valid norm $\norm{\cdot}$, allowing the user to encode problem-specific geometry and regularity. We show that this flexibility can lead to substantial performance improvements in practice.
Our main contribution is methodological: we unify and generalize a range of projection-based inference schemes under a single, computationally efficient framework that is compatible with standard Bayesian workflows. In doing so, we provide a practical alternative to truncation-based methods, especially in cases where the constraint set has zero measure or is difficult to sample from directly. Theoretical results are included to support the methodology and confirm that desirable asymptotic behavior is preserved under projection.
Our framework is broadly compatible with existing software and has minimal implementation overhead.
Related to this work is  \cite{astfalck2024generalised}, who study projected posterior beliefs for probability-free methods in Bayes linear statistics, and \cite{everink2023bayesian} who define a constrained prior that enforces a projection of the posterior mass, akin to an empirical Bayes prior.

The contributions of the paper are organized as follows. In Section~\ref{sec:posterior_projected} we define the projected posterior distribution and provide a simple illustrative example. Further, we examine some motivations for our approach: we present a decision theoretic interpretation, show that the projected posterior is a minimizer to the Wasserstein distance between any distribution defined on $\tilde{\Theta}$ and $\Pi_\Theta(\theta \mid x_{(n)})$, and demonstrate the existence of an empirical Bayes prior that leads to the projected posterior distribution. Section~\ref{sec:asymptotics} confirms the standard asymptotic properties expected of a well-behaved posterior: posterior consistency, posterior contraction, and asymptotic coverage via Bernstein--von Mises (BvM) Theorems for when $\tilde{\Theta}$ is of the same and lower topological dimension of $\Theta$. While the theoretical results are natural given existing literature, they serve to confirm the validity and robustness of the proposed methodology. In Section~\ref{sec:case_studies} we demonstrate the applicability of the projected posterior distribution to a number of case studies. Throughout the manuscript, we make certain general assumptions; Section~\ref{sec:relax} provides discussion on edge-cases, extensions and relaxations to these assumptions. Finally, we provide discussion and concluding remarks in Section~\ref{sec:discussion}. Throughout, we state a number of theorems the proofs of which are all found in Appendix~\ref{app:proofs}.

\section{The Posterior Projected Distribution} \label{sec:posterior_projected}

Before introducing the formal measure-theoretic construction, we briefly summarize the core idea. Our methodology proceeds in two simple steps: (1) fit the unconstrained Bayesian model and (2) compute its projection onto the constrained space $\tilde{\Theta}$. The first step can be done using any software or implementation of choice. Most commonly the resulting posterior will be described via a set of samples $\theta^{[1]}, \dots, \theta^{[M]} \sim \Pi_\Theta(\theta \mid x_{(n)})$. The second step is then achieved by computing the projection for each draw $\theta^{[i]}$ onto $\tilde{\Theta}$. This collection then represents samples from the \emph{projected posterior}. Geometrically, one may think of this procedure as taking the unconstrained posterior distribution and \textit{pushing} its mass onto the constrained space along shortest paths defined by the projection. The remainder of this section formalizes these ideas and provides the mathematical foundations. We rely on a measure-theoretic construction which, while introducing technical overhead, provides the requisite machinery to establish a suite of theoretical properties.

\subsection{Definition}

We begin by stating the assumptions required to define the projected posterior distribution. These ensure the well-posedness of the projection operator and guarantee the measurability and regularity conditions needed for defining the push-forward measure. Assumptions~\ref{ass:subset} and \ref{ass:map} assume that $\tilde{\Theta}$ is closed and that the projection is unique, respectively. These assumptions cover many common cases in statistical inference but are, nevertheless, relatively strong. We discuss suitable relaxations in Section~\ref{sec:relax}.

\begin{assumption} \label{ass:banach}
    The unconstrained sample space $\Theta$ is a separable Banach space $(\Theta, \norm{\cdot})$.
\end{assumption}
\begin{assumption} \label{ass:subset}
    The constrained sample space $\tilde{\Theta}$ is a non-empty and closed subset of $\Theta$.
\end{assumption}
\begin{assumption} \label{ass:map}
    There exists a projection map $\mathrm{T}_{\tilde{\Theta}}: \Theta \to \tilde{\Theta}$ such that for all $\theta \in \Theta$, the projection in Equation~\eqref{eqn:proj_operator}, below, is uniquely defined.
\end{assumption}

\begin{assumption} \label{ass:finite_var}
  For each $n$ and realized data $x_{(n)}$, the unconstrained posterior distribution $\Pi_\Theta(\theta \mid x_{(n)}) \in \mathcal{P}_2(\Theta)$, the space of probability measures with finite second moment on $\Theta$.
\end{assumption}

Consider a random sample $x_{(n)} \in \mathcal{X}$ drawn from a regular conditional distribution $\mathrm{P}(x_{(n)} \mid \theta)$ on the sample space $(\mathcal{X}, \mathcal{B}_\mathcal{X})$ with $\sigma$-algebra $\mathcal{B}_\mathcal{X}$. The parameter $\theta \in \Theta \subseteq \mathbb{R}^d$ has prior probability measure $\Pi_\Theta$ on $(\Theta, \mathcal{B}_\Theta)$. If $\mathrm{P}(x_{(n)} \mid \theta)$ is a dominated collection of measures, then there exists a density $\mathrm{p}(x_{(n)} \mid \theta)$ with respect to a $\sigma$-finite dominating measure $\mu$, typically a Radon measure, such that the map $(x_{(n)}, \theta) \mapsto \mathrm{p}(x_{(n)} \mid \theta)$ is jointly measurable. As a result, $(x_{(n)}, \theta)$ has a well-defined joint distribution on the product measure space $(\mathcal{X} \times \Theta, \mathcal{B}_\mathcal{X} \times \mathcal{B}_\Theta)$ given by $\mathrm{P}(x_{(n)} \in A, \theta \in B) = \int_B \mathrm{p}(A \mid \theta) \; \mathrm{d}\Pi_\Theta(\theta)$ for $A \in \mathcal{B}_\mathcal{X}$ and $B \in \mathcal{B}_\Theta$. According to Assumption~\ref{ass:subset}, $\tilde{\Theta} \subseteq \Theta$ is a non-empty, closed subset representing some parameter constraint. We similarly equip $\tilde{\Theta}$ with a $\sigma$-algebra $\mathcal{B}_{\tilde{\Theta}}$ and fix an arbitrary prior probability measure $\Pi_{\tilde{\Theta}}$ on $(\tilde{\Theta}, \mathcal{B}_{\tilde{\Theta}})$. Under the prior $\Pi_{\tilde{\Theta}}$, the posterior measure is
\begin{equation} \label{eqn:bayes}
    \Pi_{\tilde{\Theta}}(\tilde{B} \mid x_{(n)}) = \frac{\int_{\tilde{B}} \mathrm{p}(x_{(n)} \mid \theta) \; \mathrm{d}\Pi_{\tilde{\Theta}}(\theta)}{\int_{\tilde{\Theta}} \mathrm{p}(x_{(n)} \mid \theta) \; \mathrm{d}\Pi_{\tilde{\Theta}}(\theta)}
\end{equation}
for all $\tilde{B} \in \mathcal{B}_{\tilde{\Theta}}$. Since $\Theta$ is a Polish space (Assumption~\ref{ass:banach}), by Alexandrov's Theorem we have $\mathcal{B}_{\tilde{\Theta}} = \{B \cap \tilde{\Theta} : B \in \mathcal{B}_\Theta\}$ ensuring \eqref{eqn:bayes} is well defined \citep{ghosal2017fundamentals}. When $\tilde{\Theta} = \Theta$, \eqref{eqn:bayes} induces the posterior distribution $\Pi_{\Theta}(\theta \mid x_{(n)})$, which we refer to as the \textit{unconstrained} posterior distribution.

To define the projected posterior distribution, we exploit the geometric properties of $(\Theta, \norm{\cdot})$ to project the unconstrained posterior measure into $\tilde{\Theta}$. Given the associated norm $\norm{\cdot}$ of the Banach space $\Theta$, define a distance between a point $\theta \in \Theta$ and the subset $\tilde{\Theta}$ as $\mathrm{dist}(\theta, \tilde{\Theta}) = \mathrm{inf}\{\lVert\theta - \tilde{\theta}\rVert : \tilde{\theta} \in \tilde{\Theta}$\}. We are now in a position to formally define the map in Assumption~\ref{ass:map} as
\begin{equation} \label{eqn:proj_operator}
    \mathrm{T}_{\tilde{\Theta}}(\theta) = \left\{ \tilde{\theta} \in \tilde{\Theta} : \lVert\theta - \tilde{\theta}\rVert = \mathrm{dist}(\theta, \tilde{\Theta}) \right\},
\end{equation}
where, as per Assumption~\ref{ass:map}, we assume $\mathrm{T}_{\tilde{\Theta}}$ is unique for all $\theta \in \Theta$. There are many sufficient conditions on $\tilde{\Theta}$ that satisfy Assumption~\ref{ass:map}. The most common of these is to assume that $\tilde{\Theta}$ is a closed convex subset and that the norm $\norm{\cdot}$ is induced by an inner product $\langle \cdot, \cdot \rangle$ so that the Banach space $(\Theta, \norm{\cdot})$ is induced by a Hilbert space $(\Theta, \langle \cdot, \cdot \rangle)$. The Hilbert Projection Theorem then ensures Assumption~\ref{ass:map} holds. Another well known result that satisfies Assumption~\ref{ass:map} is when $\tilde{\Theta}$ is a Stiefel manifold endowed with the trace inner product \citep{absil2012projection}. We provide examples of both of these conditions in the case studies in Section~\ref{sec:case_studies}.

Define the inverse image $\mathrm{T}_{\tilde{\Theta}}^{-1} (\tilde{B}) = \{\theta \in \Theta : \mathrm{T}_{\tilde{\Theta}} (\theta) \in \tilde{B}\}$. Due to Assumption~\ref{ass:map}, $\mathrm{T}_{\tilde{\Theta}}$ is measurable as $\mathrm{T}_{\tilde{\Theta}}^{-1} (\tilde{B}) \in \mathcal{B}_\Theta$ for all $\tilde{B} \in \mathcal{B}_{\tilde{\Theta}}$. Thus, provided with the unconstrained posterior measure $\Pi_\Theta(B \mid x_{(n)})$ on $(\Theta, \mathcal{B}_\Theta)$, $\mathrm{T}_{\tilde{\Theta}}$ induces a push-forward measure $\tilde{\Pi}_{\tilde{\Theta}}(\tilde{B} \mid x_{(n)})$ on $(\tilde{\Theta}, \mathcal{B}_{\tilde{\Theta}})$, such that for any $\tilde{B} \in \mathcal{B}_{\tilde{\Theta}}$,
\begin{equation}\label{eqn:proj_posterior}
    \tilde{\Pi}_{\tilde{\Theta}}(\tilde{B} \mid x_{(n)}) = \Pi_{\Theta}(\mathrm{T}_{\tilde{\Theta}}^{-1} (\tilde{B}) \mid x_{(n)}).
\end{equation}
We define this distribution, $\tilde{\Pi}_{\tilde{\Theta}}(\theta \mid x_{(n)})$, as the \textit{projected} posterior distribution. Note that \eqref{eqn:proj_posterior} is strictly with respect to $\norm{\cdot}$; however, for simplicity, we have suppressed this in the notation so that the dependency is implicit.

\subsection{A simple example} \label{sec:simple_example}

We illustrate the construction with a simple example. Consider data $x_{(n)} = (x_1, \dots, x_n)$ sampled from  $x_i \mid \theta \overset{\mathrm{iid}}{\sim} \mathcal{N}(x; \theta, 1)$ for $i \in \{1, \dots, n\}$. Say we have the unconstrained prior belief $\theta \sim \mathcal{N}(\mu_0, n_0^{-1})$; this yields the unconstrained posterior distribution $\Pi_\Theta(\theta \mid x_{(n)}) = \mathcal{N}(\mu_n, \sigma_n^2)$ for $\theta \in \Theta = \mathbb{R}$, where $\sigma_n^2 = (n + n_0)^{-1}$, $\mu_n = \sigma_n^2(n \bar{x} + n_0 \mu_0)$ and $\bar{x} = n^{-1} \sum_i x_i$. Say we believe $\theta \in \tilde{\Theta} = \mathbb{R}^+$ is the non-negative real numbers, noting that despite the open and infinite upper bound, $\tilde{\Theta}$ is a closed set, as the complement $\tilde{\Theta}^c$ is open. Denote by $\mathcal{TN}_{[a,b]}(\cdot)$ the truncated normal distribution truncated on the interval $[a,b]$; the truncated posterior is estimated by specifying the truncated prior $\mathcal{TN}_{[0,\infty)}(\mu_0, n_0^{-1}) \propto \mathcal{N}(\mu_0, n_0^{-1}) \mathds{1}\{\theta \in \mathbb{R}^+\}$ so that $\Pi_{\tilde{\Theta}}(\theta \mid x_{(n)}) = \mathcal{TN}_{[0,\infty)}(\mu_n, \sigma_n^2)$. This posterior distribution assigns zero probability measure to the boundary $\{0\}$ and has expectation and variance
\begin{equation*}
    \mathbb{E}_{\Pi_{\tilde{\Theta}}}(\theta \mid x_{(n)}) = \mu_n + \frac{\varphi(\alpha)}{1 - \Phi(\alpha)} \sigma_n, \quad \mathrm{var}_{\Pi_{\tilde{\Theta}}}(\theta \mid x_{(n)}) = \sigma_n^2\left[1 + \left(\frac{\varphi(\alpha)}{1 - \Phi(\alpha)}\right)^2\right]
\end{equation*}
where $\alpha = -\mu_n/\sigma_n$, $\varphi(\cdot)$ and $\Phi(\cdot)$ denote the standard normal probability density function and cumulative density function, respectively, and the expectation and variance are with respect to the truncated posterior measure $\Pi_{\tilde{\Theta}}$.

We calculate the projected posterior distribution by projecting the unconstrained posterior distribution into $\tilde{\Theta} = \mathbb{R}^+$ via the map defined in \eqref{eqn:proj_operator}. To do so we would typically be required to specify $\norm{\cdot}$, the associated norm of the Banach space $(\Theta, \norm{\cdot})$, although in this instance the problem is invariant to the specification of $\norm{\cdot}$ as $\Theta \in \mathbb{R}$. Accordingly, we leave discussion of the selection of $\norm{\cdot}$ to later sections of the manuscript. The projected posterior is
\begin{equation} \label{eqn:normal_proj_posterior}
    \tilde{\Pi}_{\tilde{\Theta}}(\theta \mid x_{(n)}) = \Phi(\alpha) \delta(0) + \{1 - \Phi(\alpha)\} \mathcal{TN}_{[0,\infty)}(\mu_n, \sigma_n^2)
\end{equation}
where $\delta(0)$ is the Dirac delta measure at zero. The projected posterior expectation and variance are
\begin{equation*}
\begin{aligned}
  \mathbb{E}_{\tilde{\Pi}_{\tilde{\Theta}}}(\theta \mid x_{(n)}) 
&= \{1 - \Phi(\alpha)\}\,\mathbb{E}_{\Pi_{\tilde{\Theta}}}(\theta \mid x_{(n)}), \\
\mathrm{Var}_{\tilde{\Pi}_{\tilde{\Theta}}}(\theta \mid x_{(n)})
&= \mathbb{E}_{\tilde{\Pi}_{\tilde{\Theta}}}(\theta^2 \mid x_{(n)})
- \left\{\mathbb{E}_{\tilde{\Pi}_{\tilde{\Theta}}}(\theta \mid x_{(n)})\right\}^2 \\
&= \{1 - \Phi(\alpha)\}\,\mathrm{Var}_{\Pi_{\tilde{\Theta}}}(\theta \mid x_{(n)})
\;+\;
\Phi(\alpha)\{1 - \Phi(\alpha)\}
\left\{\mathbb{E}_{\Pi_{\tilde{\Theta}}}(\theta \mid x_{(n)})\right\}^{2}.
\end{aligned}
\end{equation*}
Due to the point mass at zero, the projected posterior distribution in \eqref{eqn:normal_proj_posterior} does not have a density with respect to the Lebesgue measure as with the unconstrained and truncated posteriors. However, it has the density
\begin{equation} \label{eqn:normal_density}
    \tilde{\pi}_{\tilde{\Theta}}(\theta \mid x_{(n)}) = \Phi(\alpha) \mathds{1}\{\theta = 0\} + \left\{ 1 - \Phi(\alpha)\right\} \varphi_{[0, \infty)}(\theta; \mu_n, \sigma_n^2),
\end{equation}
with respect to the Radon measure
\begin{equation} \label{eqn:normal_ref_measure}
  \lambda_{[0, \infty)}(\tilde{B}) = \mathds{1}\{0 \in \tilde{B}\} + \mu_\mathbb{R}([0, 
  \infty) \cap \tilde{B})
\end{equation}
where $\mu_\mathbb{R}$ denotes the Lebesgue measure in $\mathbb{R}$, and $\varphi_{[0, \infty)}(\theta; \mu_n, \sigma_n^2)$ denotes the density of the $[0, \infty)$-truncated normal distribution.

\subsection{Motivation for the projected posterior} \label{sec:motivations}

We now present some properties of the projected posterior distribution that motivate its use. The measure-theoretic construction above is sufficient to define and implement the projected posterior; the following interpretations further clarify its behavior and motivate its use.

\subsubsection{Decision theoretic interpretation}

Bayesian decision theory aims to minimize the expected loss of a decision with respect to the posterior distribution. Define the decision $\xi$ and loss function $\mathrm{L}(\theta, \xi)$. Strictly, $\xi$ is dependent on the observed data $x_{(n)}$ and prior beliefs $\Pi_{\Theta}(\theta)$ but we do not include these dependencies in the notation for simplicity. The Bayes decision rule is the $\xi$ that solves
\begin{equation*}
  \xi^* = \underset{\xi}{\arg \min} \int_{\Theta} \mathrm{L}(\theta, \xi) \; \mathrm{d}\Pi_{\Theta}(\theta \mid x_{(n)}).
\end{equation*}
For instance, setting $\mathrm{L}(\theta, \xi) = (\xi - \theta)^2$ as squared loss leads to the Bayes decision rule being equal to the posterior mean, $\xi^* = \mathbb{E}[\theta \mid x_{(n)}]$. The value of the expected loss given the Bayes decision rule is known as Bayes risk,
\begin{equation*}
  \mathrm{R}(\xi^*) = \int_{\Theta} \mathrm{L}(\theta, \xi^*) \; \mathrm{d}\Pi_{\Theta}(\theta \mid x_{(n)}).
\end{equation*}
We state the following theorem, in regards to the Bayes decision rule of the projected posterior distribution.

\begin{theorem} \label{the:decision}
  Given a loss function $\mathrm{L}(\theta, \xi)$ with decision $\xi$, the Bayesian decision rule under $\tilde{\Pi}_{\tilde{\Theta}}(\theta \mid x_{(n)})$ is equal to
  \begin{equation*}
    \xi^* = \underset{\xi}{\arg \min} \int_\Theta \mathrm{L}(\theta, \xi) \; \mathrm{d}\tilde{\Pi}_{\tilde{\Theta}}(\theta \mid x_{(n)}) = \underset{\xi}{\arg \min} \int_\Theta \mathrm{L}(\mathrm{T}_{\tilde{\Theta}}(\theta), \xi) \; \mathrm{d}\Pi_{\Theta}(\theta \mid x_{(n)}).
  \end{equation*}
\end{theorem}
The proof is provided in Appendix~\ref{app:proofs}.

Theorem~\ref{the:decision} states that the Bayesian decision rule under the projected posterior distribution is equal to the Bayesian decision rule of $\mathrm{L}(\mathrm{T}_{\tilde{\Theta}}(\theta), \xi)$ under the unconstrained posterior distribution. Thus for Bayesian decision theory, the action of projecting alters the loss function from $\mathrm{L}(\theta, \xi)$ to $\mathrm{L}(\mathrm{T}_{\tilde{\Theta}}(\theta), \xi)$ for $\theta \in \Theta$.  A useful consequence of Theorem~\ref{the:decision} is that projection provides a lower bound for the Bayes risk for a broad classes of distance-based loss functions and projections, formalized in Corollary~\ref{cor:proj_bayes_risk}.

\begin{corollary}\label{cor:proj_bayes_risk}
Suppose the loss is of the form $\mathrm{L}(\theta,\xi)=\ell\!\left(\|\theta-\xi\|\right)$ where $\ell:[0,\infty)\to\mathbb{R}$ is non-decreasing. Assume $\mathrm{T}_{\tilde{\Theta}}$ has Lipschitz constant $K \leq 1$ such that $\norm{\mathrm{T}_{\tilde{\Theta}}(\theta) - \mathrm{T}_{\tilde{\Theta}}(\theta')} \leq \norm{\theta - \theta'}$. The Bayes risk under the projected posterior distribution lower bounds the standard Bayes risk such that
\[
\int_{\tilde{\Theta}}\mathrm{L}(\theta,\xi)\,d\tilde{\Pi}_{\tilde{\Theta}}(\theta\mid x_{(n)})
\le
\int_{\Theta}\mathrm{L}(\theta,\xi)\,d\Pi_{\Theta}(\theta\mid x_{(n)}),
\]
and consequently,
\[
\underset{\xi\in\Xi}{\min}\int_{\tilde{\Theta}}\mathrm{L}(\theta,\xi)\,d\tilde{\Pi}_{\tilde{\Theta}}(\theta\mid x_{(n)})
\le
\underset{\xi\in\Xi}{\min}\int_{\Theta}\mathrm{L}(\theta,\xi)\,d\Pi_{\Theta}(\theta\mid x_{(n)})
\]
for the entire decision space $\xi \in \Xi\subseteq\tilde{\Theta}$.
\end{corollary}

Proof is provided in Appendix~\ref{app:proofs}. We note that the condition $K \leq 1$ is satisfied when $\tilde{\Theta}$ is a closed convex subset of a Hilbert space, since the associated metric projection is non-expansive.

\subsubsection{The Wasserstein distance}

Denote by $\mathcal{P}_2(\tilde{\Theta})$ the set of all probability measures on $(\tilde{\Theta}, \mathcal{B}_{\tilde{\Theta}})$ with finite second-order moments. Further recall that any probability measure on $(\tilde{\Theta}, \mathcal{B}_{\tilde{\Theta}})$ is a valid probability measure on $(\Theta, \mathcal{B}_{\Theta})$. The Wasserstein-$2$ distance between any two probability measures $\mu, \nu \in \mathcal{P}_2(\Theta)$ and with respect to the norm $\norm{\cdot}$ is defined as
\begin{equation} \label{eqn:wasserstein}
    \mathcal{W}_2(\mu, \nu) = \underset{\gamma \in \Gamma(\mu, \nu)}{\inf} \left\{\int_{\Theta \times \Theta} \norm{\theta - \theta'}^2 \; \mathrm{d} \gamma(\theta, \theta')\right\}^{1/2}
\end{equation}
where $\Gamma(\mu, \nu)$ is the family of all probability measures on the product space $\Theta \times \Theta$ with marginals $\mu$ and $\nu$. The Wasserstein distance is the natural way to measure the distance between two arbitrary distributions defined on a Banach space $(\Theta, \norm{\cdot})$. Commonly, \eqref{eqn:wasserstein} is defined by a metric $d(\theta, \theta')$; however, given Assumption~\ref{ass:banach}, here we find it to be more natural to define \eqref{eqn:wasserstein} via the norm $\norm{\theta - \theta'}$. If we set $\mu$ in \eqref{eqn:wasserstein} to be the unconstrained posterior we obtain the following theorem.

\begin{theorem} \label{the:wasserstein}
  Given Assumptions~\ref{ass:banach}--\ref{ass:finite_var}, the projected posterior distribution is the element in $\mathcal{P}_2(\tilde{\Theta})$ that minimizes the Wasserstein-2 distance from the unconstrained posterior distribution,
    \begin{equation*}
        \tilde{\Pi}_{\tilde{\Theta}}(\theta \mid x_{(n)}) = \underset{\nu \in \mathcal{P}_2(\tilde{\Theta})}{\arg\min} \left\{\mathcal{W}_2(\nu, \Pi_\Theta(\theta \mid x_{(n)})) \right\}
    \end{equation*}
  with respect to the norm $\norm{\cdot}$ that defines $(\Theta, \norm{\cdot})$.
\end{theorem}
  The proof is provided in Appendix~\ref{app:proofs}.

Theorem~\ref{the:wasserstein} implies that with respect to $\mathcal{W}_2(\cdot, \cdot)$, the projected posterior distribution is the closest distribution to the unconstrained posterior of all distributions in $\mathcal{P}_2(\tilde{\Theta})$. Thus, when the belief specifications simultaneously yield a belief in the unconstrained posterior distribution and a belief that the posterior measure must lie in $\tilde{\Theta}$, the projected posterior distribution is the optimal choice of distribution.

\subsubsection{An Empirical Bayes Prior}

Empirical Bayes techniques estimate the prior distribution of a statistical model from the data \citep{maritz2018empirical}. Consider a prior distribution $\Pi_\Theta(\theta)$, which combined with a likelihood and observations $x_{(n)}$ yields the unconstrained posterior $\Pi_\Theta(\theta \mid x_{(n)}, \Pi_\Theta(\theta))$. Here, we have been explicit about the dependency of the unconstrained posterior on the prior. We show in this section that an empirical Bayes prior $\tilde{\Pi}_{\tilde{\Theta}}(\theta)$ exists such that the unconstrained posterior calculated under $\tilde{\Pi}_{\tilde{\Theta}}(\theta)$, $\Pi_\Theta(\theta \mid x_{(n)}, \tilde{\Pi}_{\tilde{\Theta}}(\theta))$ is the same as the projected posterior calculated under $\Pi_\Theta(\theta)$, $\tilde{\Pi}_{\tilde{\Theta}}(\theta \mid x_{(n)}, \Pi_\Theta(\theta))$, almost everywhere. Mathematically, this implies for any $\tilde{B} \in \mathcal{B}_{\tilde{\Theta}}$, $\Pi_\Theta(\tilde{B} \mid x_{(n)}, \tilde{\Pi}_{\tilde{\Theta}}(\tilde{B})) = \tilde{\Pi}_{\tilde{\Theta}}(\tilde{B} \mid x_{(n)}, \Pi_\Theta(\tilde{B}))$. First, we establish the existence of a projected posterior density in Lemma~\ref{lem:radon_nikodym}.

\begin{lemma} \label{lem:radon_nikodym}
  The projected posterior distribution $\tilde{\Pi}_{\tilde{\Theta}}(\theta \mid x_{(n)})$ has a density $\tilde{\pi}_{\tilde{\Theta}}(\theta \mid x_{(n)})$ given by its Radon-Nikodym derivative with respect to a $\sigma$-finite reference measure $\tilde{\mu}$ on $\tilde{\Theta}$.
\end{lemma}
  The proof is provided in Appendix~\ref{app:proofs}.

Lemma~\ref{lem:radon_nikodym} allows us to establish a data-dependent prior distribution that leads to the projected posterior density via a traditional Bayes update.

\begin{theorem} \label{the:projected_prior}
  As in Lemma~\ref{lem:radon_nikodym}, denote by $\tilde{\pi}_{\tilde{\Theta}}(\theta \mid x_{(n)})$ the density of the projected posterior distribution given by its Radon-Nikodym derivative with respect to a $\sigma$-finite reference measure $\tilde{\mu}$ on $\tilde{\Theta}$. The unconstrained posterior calculated from the empirical prior density 
  \begin{equation}\label{eqn:projected_prior}
    \tilde{\pi}_{\tilde{\Theta}}^{\mathrm{EB}}(\theta; x_{(n)}) \coloneq \mathrm{p}(x_{(n)} \mid \theta)^{-1} \tilde{\pi}_{\tilde{\Theta}}(\theta \mid x_{(n)})\left[\int_{\tilde{\Theta}} \mathrm{p}(x_{(n)} \mid \theta)^{-1} \tilde{\pi}_{\tilde{\Theta}}(\theta \mid x_{(n)}) \; \mathrm{d}\tilde{\mu}(\theta)\right]^{-1}, \theta \in \tilde{\Theta}
  \end{equation}
  is the same as the projected posterior distribution almost everywhere, provided that $p(x_{(n)} \mid \theta) > 0$ $\tilde{\mu}$-almost everywhere on the support of $\tilde{\pi}_{\tilde{\Theta}}(\cdot \mid x_{(n)})$ and that the normalizing constant in \eqref{eqn:projected_prior} is finite and non-zero.
\end{theorem}
  The proof is provided in Appendix~\ref{app:proofs}.

The result of Theorem~\ref{the:projected_prior} is that the projected posterior may be viewed as a certain type of empirical Bayes posterior. For certain simple cases \eqref{eqn:projected_prior} is analytically tractable, for instance, in the example in Section~\ref{sec:simple_example}. We conjecture that a functional form for \eqref{eqn:projected_prior} is available when $\tilde{\Theta}$ is a finite convex polytope, $\mathrm{p}(x_{(n)} \mid \theta)$ is Gaussian and $\tilde{\pi}_{\tilde{\Theta}}(\theta \mid x_{(n)})$ is a sum of Gaussians of dimension less than or equal to $\mathrm{dim}(\Theta)$, including the Dirac measure. Such $\tilde{\pi}_{\tilde{\Theta}}^{\mathrm{EB}}(\theta; x_{(n)})$  quickly becomes computationally burdensome, and for the more general case, would need to be calculated numerically. Studying when the empirical prior in \eqref{eqn:projected_prior} is independent of the data, i.e. $\tilde{\pi}_{\tilde{\Theta}}^{\mathrm{EB}}(\theta; x_{(n)}) = q(\theta)$ leads to Corollary~\ref{cor:emp_bayes}.

\begin{corollary} \label{cor:emp_bayes}
There exists a density $q$ on $\tilde\Theta$ such that $\tilde{\pi}_{\tilde{\Theta}}^{\mathrm{EB}}(\theta; x_{(n)}) = q(\theta)$ for all $x_{(n)} \in \mathcal{X}$ and $\tilde\mu$-almost every $\theta \in \tilde{\Theta}$
if and only if there exists a density $q$ on $\tilde\Theta$ such that, for all $x_{(n)}$,
\[
\tilde{\pi}_{\tilde{\Theta}}(\theta\mid x_{(n)}) =
\frac{\mathrm{p}(x_{(n)}\mid \theta)\,q(\theta)}
{\int_{\tilde{\Theta}} \mathrm{p}(x_{(n)}\mid u)\,q(u)\,\mathrm d\tilde\mu(u)}
\]
for $\tilde\mu$-almost every $\theta \in \tilde{\Theta}$.
\end{corollary}
Corollary~\ref{cor:emp_bayes} implies the only way the empirical Bayes prior in \eqref{eqn:projected_prior} can be data-free is if the projected posterior is already a Bayesian posterior under the same likelihood. The implication is that the projected posterior is a strictly broader class of distribution than the Bayesian posterior and thus provides more flexible inference.

\subsection{Choosing the norm} \label{sec:norm}

So far we have not explicitly discussed the appropriate selection of the norm in the definition of the projected posterior. The theory and definition exist for any norm, and there are certainly situations in which the choice of one norm over another may be sensible for reasons of tractability or convenience. Indeed, this flexibility is one of the advantages of our theory. There are two natural defaults: the Euclidean norm $\norm{\cdot}_2$ and the Mahalanobis norm $\norm{\cdot}_\Sigma$. For many constraints, projection under the Euclidean norm is analytic. As the driving motivation for our theory is one of principled convenience, this may provide a strong argument for its selection. However, the Euclidean norm does not respect any geometrical properties of the posterior distribution. In contrast, the Mahalanobis norm measures distance relative to posterior variance, noting that covariance forms a valid inner product that may define a norm distance. In particular, it scales directions according to posterior uncertainty, so that deviations in highly uncertain directions are penalized less heavily than those in well-identified directions. In regular parametric models where the posterior is asymptotically Gaussian, this geometry arises naturally from local asymptotic normality and the norm projection will align with the Wasserstein geodesics in probability space.

Assume, as per Assumption~\ref{ass:finite_var}, that the unconstrained posterior has a finite second moment, and further assume $\Theta \subseteq \mathbb{R}^d$ is finite dimensional so that $d \in \mathbb{N}$. Denote by $\Sigma$ the posterior variance matrix with $(i,j)$th elements $\Sigma_{i,j} = \mathrm{cov}[\theta_i, \theta_j \mid x_{(n)}]$ where $\theta_i$ is the random quantity represented by the $i$th dimension of $\Theta$. Further denote by $\lambda_i$ and $\varphi_i$ the $i$th eigenvalue and eigenvector of $\Sigma$, respectively. Covariance naturally defines a reproducing kernel Hilbert space (RKHS) with the inner product $\langle \cdot, \cdot \rangle_\Sigma$. From Mercer's Theorem, define the RKHS defined by $\Sigma$ as
\begin{equation} \label{eqn:RKHS}
  \mathcal{H}_\Sigma \coloneq \left\{ \theta \in \Theta : \sum_{i = 1}^d \frac{\langle \theta, \varphi_i \rangle^2_2}{\lambda_i} < \infty \right\}
\end{equation}
where $\langle \cdot, \cdot \rangle_2$ is the standard Euclidean inner product. The inner product and squared norm of $\mathcal{H}_\Sigma$ are respectively
\begin{equation}\label{eqn:RKHS_ip_and_norm}
  \langle \theta, \theta' \rangle_\Sigma = \sum_{i = 1}^d \frac{\langle \theta, \varphi_i \rangle_2 \langle \theta', \varphi_i \rangle_2}{\lambda_i}, \quad \text{and} \quad \norm{\theta}^2_\Sigma = \sum_{i = 1}^d \frac{\langle \theta, \varphi_i \rangle^2_2}{\lambda_i}.
\end{equation}
Thus, in $\mathcal{H}_\Sigma$, the norm that defines the projection operator in \eqref{eqn:proj_operator} is given by the Mahalanobis distance,
\begin{equation} \label{eqn:mahalnobis}
  \norm{\theta - \theta'}_\Sigma^2 = (\theta - \theta')^\intercal \Sigma^{-1} (\theta - \theta').
\end{equation}
As the norm is defined from an RKHS, the Hilbert Projection Theorem states that a sufficient condition that satisfies the uniqueness requirement in Assumption~\ref{ass:map} is that $\tilde{\Theta}$ is convex. 

In the case studies below, we provide different use cases that motivate different choices of norm. Calculation of the Mahalanobis norm presupposes the knowledge of the value of $\Sigma$. This makes the Mahalanobis norm a sensible choice for projecting Gaussian posteriors (see Case Studies 4 and 5) or when there is an approximate asymptotic posterior form that may be used as a surrogate (see Case Study 1). It is common for $\Sigma$ to be a function of $\theta$, or for the use of the Mahalanobis norm to break analytical tractability that exists under an Euclidean norm, and so choice of Euclidean norm may be preferable (see Case Study 2 and 3).

\subsection{Implementation and Computational Cost}

Posterior projection is a post hoc operation that requires no modification to the underlying Bayesian model. For simple problems, such as that in Section~\ref{sec:simple_example}, the posterior projection can be calculated exactly and described by an analytical measure. For problems of realistic scientific complexity, the unconstrained posterior measure will be described as a sequence of samples from a MCMC scheme (or similar): $\{\theta^{[1]},\dots,\theta^{[M]}\}$. The projected posterior is simply described as the projection onto the constraint set via \[
\tilde{\theta}^{[i]} = \underset{\tilde{\theta}\in\tilde{\Theta}}{\arg\min}
\|\theta^{[i]} - \tilde{\theta}\|.
\]
The projected samples $\{\tilde{\theta}^{[i]}\}_{i=1}^M$
constitute draws from the projected posterior. The proof follows 
by equating a projected draw $\tilde{\theta}^{[i]}$ with $\tilde{B}$ in the definition in \eqref{eqn:proj_posterior}.

The per sample computational cost of projection depends on both the dimension $d = \mathrm{dim}(\Theta)$ and the geometry of the constraint set $\tilde{\Theta}$. For the Euclidean norm, many common constraint sets admit closed-form or linear-time projections (e.g., box constraints, simplex constraints, order constraints, isotonic
regression). Under the Mahalanobis norm, forming the metric $\norm{\cdot}_\Sigma$ requires a Cholesky factorization of $\Sigma$, which in general scales as $\mathcal{O}(d^3)$. This is a one-time cost and can be reduced substantially when 
$\Sigma$ is sparse or structured, for example via Vecchia or other sparse approximations \citep[e.g.][]{katzfuss2021general}. Each projection then requires solving a quadratic program of the form
$
\min_{\tilde{\theta} \in \tilde{\Theta}} 
(\theta - \tilde{\theta})^\intercal 
\Sigma^{-1} 
(\theta - \tilde{\theta}),
$
whose complexity depends on the structure of $\tilde{\Theta}$. For instance, for convex polyhedral constraint sets defined by $k$ linear inequalities, interior-point methods typically scale as $\mathcal{O}(d^3 + d^2 k + k^3)$ per projection, while first-order methods reduce this to approximately $\mathcal{O}(d^2)$ per iteration \citep{nocedal2006numerical}. Consequently, for $M$ posterior samples, the overall projection cost 
scales as $\mathcal{O}(d^3) + M \cdot C_{\mathrm{proj}}(d,k)$, where $C_{\mathrm{proj}}(d,k)$ reflects the geometry of $\tilde{\Theta}$ and the optimization method employed. Importantly, the projections are embarrassingly parallel across samples, so parallel architectures can substantially reduce wall-clock time.

\section{Asymptotic Results} \label{sec:asymptotics}

This section presents asymptotic results that validate the use of the projected posterior. These results establish posterior consistency, the posterior contraction rate, and two BvM Theorems for when $\tilde{\Theta}$ is of the same and lower topological dimension as $\Theta$. These results show that under well-specified constraints, the projected posterior inherits the asymptotic behavior of the unconstrained posterior. These findings should be interpreted as a consistency check for the proposed method, not as a statement that it dominates existing alternatives such as truncation or transformation. Indeed, we conjecture that similar results will also apply to the truncated posterior when $\mathrm{dim}(\tilde{\Theta}) = \mathrm{dim}(\Theta)$.

\subsection{Posterior consistency} \label{sec:consistency}

For simplicity, here we assume the existence of a true parameter $\theta_0$. Many subjective Bayesians agree with such an assumption; to obtain similar proofs more in line with the subjectivist point of view, posterior consistency could be defined as convergence of posterior predictive distributions under differing prior beliefs defined on the same support \citep[see][]{diaconis1986consistency}, with $\theta_0$ the center of the smallest ball that contains $0 < c < 1$ of the posterior mass as $n \rightarrow \infty$. Given $\theta_0$, we define posterior consistency similarly to \cite{ghosal2017fundamentals}, and say that the posterior distribution $\Pi_\Theta(\theta \mid x_{(n)})$ is weakly consistent at $\theta_0$ if $\Pi_\Theta(\theta : \norm{\theta - \theta_0} > \epsilon \mid x_{(n)}) \rightarrow 0$ in $\mathrm{P}(x_{(n)} \mid \theta_0)$-probability as $n \rightarrow \infty$ for all $\epsilon > 0$. We now state Theorem~\ref{the:consistency}, which provides conditions on consistency for the projected posterior.

\begin{theorem} \label{the:consistency}
  If Assumptions~\ref{ass:banach}--\ref{ass:map} are met, $\theta_0 \in \tilde{\Theta}$, and the unconstrained posterior distribution $\Pi_\Theta(\theta \mid x_{(n)})$ is weakly consistent at $\theta_0$, so that $\Pi_\Theta(\theta : \norm{\theta - \theta_0} > \epsilon \mid x_{(n)}) \rightarrow 0$ in $\mathrm{P}(x_{(n)} \mid \theta_0)$-probability as $n \rightarrow \infty$ for all $\epsilon > 0$, then the projected posterior distribution $\tilde{\Pi}_{\tilde{\Theta}}(\theta \mid x_{(n)})$ is weakly consistent at $\theta_0$. 
\end{theorem}
  The proof is provided in Appendix~\ref{app:proofs}.

If $\theta_0 \notin \tilde{\Theta}$ then $\tilde{\Pi}_{\tilde{\Theta}}(\theta \mid x_{(n)})$ is not consistent as $\mathrm{T}_{\tilde{\Theta}}(\theta_0) \neq \theta_0$. This property is not specific to the projected posterior distribution, but is shared with all distributions with unit measure on $\tilde{\Theta}$ including the truncated posterior distributions. This result is reassuring, as the posterior distribution is absolutely continuous with respect to the prior. Thus, when $\theta_0 \notin \tilde{\Theta}$ the failure of the projected posterior to achieve consistency is a failure of the prior belief specification.

\subsection{Posterior contraction}

Posterior contraction describes the rate at which the posterior distribution concentrates about $\theta_0$. We say that a posterior distribution $\Pi(\theta \mid x_{(n)})$ has a contraction rate $\epsilon_n$ with respect to a semimetric $d$ if
\begin{equation*}
  \Pi_\Theta(\theta:d(\theta, \theta_0) > M_n \epsilon_n \mid x_{(n)}) \rightarrow 0
\end{equation*}
in $\mathrm{P}(x_{(n)} \mid \theta_0)$-probability for all $M_n \rightarrow \infty$. In addition, we make the following assumption on $\Pi_\Theta(\theta \mid x_{(n)})$.
\begin{assumption} \label{ass:bi_lipshitz}
  The unconstrained posterior distribution contracts at rate $\epsilon_n$ with respect to a bi-Lipschitz semimetric $d$ on $(\Theta, \norm{\cdot})$ such that there exists a constant $c \geq 1$ where $c^{-1} \norm{\theta - \theta'} \leq d(\theta, \theta') \leq c \norm{\theta - \theta'}$ for all $\theta, \theta' \in \Theta$.  
\end{assumption}
This ensures that contraction with respect to $d(\theta,\theta_0)$ is equivalent, up to constant factors, to contraction in the norm $\|\theta-\theta_0\|$. Assumption~\ref{ass:bi_lipshitz} is mild in standard finite-dimensional parametric settings: in models admitting local asymptotic normality the assumption is automatically satisfied. The condition primarily excludes severely non-regular or non-identifiable models in which the posterior geometry is not locally equivalent to a fixed norm. We may now state the following theorem on the contraction rate of $\tilde{\Pi}_{\tilde{\Theta}}(\theta \mid x_{(n)})$.
\begin{theorem} \label{the:contraction}
  If $\theta_0 \in \Theta$, and Assumptions~\ref{ass:banach}--\ref{ass:map} and \ref{ass:bi_lipshitz} are met, then the projected posterior distribution $\tilde{\Pi}_{\tilde{\Theta}}(\theta \mid x_{(n)})$ satisfies
  \begin{equation*}
    \tilde{\Pi}_{\tilde{\Theta}}(\theta:d(\theta,\theta_0) > 2c^2 M_n \epsilon_n \mid x_{(n)}) \rightarrow 0
  \end{equation*}
  in probability for $c \geq 1$ and every $M_n \rightarrow \infty$. Thus, the projected posterior achieves a contraction rate at least that of the unconstrained posterior distribution.
\end{theorem}
 The proof is provided in Appendix~\ref{app:proofs}.

\subsection{Bernstein--von Mises Theorem} \label{sec:bernstein}

The BvM Theorem describes the asymptotic behavior of the posterior distribution, stating that under suitable regularity conditions, the posterior converges in total variation to a normal distribution centered at the true parameter $\theta_0$ with covariance given by the inverse Fisher information matrix. In the following, let $\hat{\theta}_n$ denote the maximum likelihood estimator of $\theta$, $\mathcal{I}(\theta_0)$ the Fisher information matrix at $\theta_0$, and $\norm{\cdot}_{\mathrm{TV}}$ the total variation distance.

\begin{assumption} \label{ass:bernstein}
  The unconstrained posterior distribution satisfies the BvM Theorem
  \begin{equation*}
    \mathbb{E}_{\mathrm{P}(x_{(n)} \mid \theta_0)}\norm{\Pi_\Theta(\theta \mid x_{(n)}) - \mathcal{N}\left(\hat{\theta}_n, \frac{1}{n} \mathcal{I}^{-1}(\theta_0)\right)}_\mathrm{TV} \rightarrow 0 \quad \text{as } n \rightarrow \infty
  \end{equation*}
  and the sequence $\hat{\theta}_n$ is asymptotically normal $\sqrt n(\hat{\theta}_n-\theta_0) \Rightarrow \mathcal N(0,\mathcal I^{-1}(\theta_0))$.
\end{assumption}

Sufficient conditions for Assumption~\ref{ass:bernstein} to hold can be found in Section~2.3 of \cite{bochkina2019bernstein}. We now show that, under suitable conditions, the projected posterior also satisfies the BvM Theorem.

\begin{theorem} \label{the:bernstein}
  Suppose $\tilde{\Theta}$ has non-empty interior in $\mathbb{R}^d$, $\theta_0 \in \tilde{\Theta}^\circ$ is an interior point of $\tilde{\Theta}$, and that Assumptions~\ref{ass:banach}--\ref{ass:map} and \ref{ass:bernstein} hold. Then the projected posterior distribution $\tilde{\Pi}_{\tilde{\Theta}}(\theta \mid x_{(n)})$ satisfies
  \begin{equation*}
    \mathbb{E}_{\mathrm{P}(x_{(n)} \mid \theta_0)}\norm{\tilde{\Pi}_{\tilde{\Theta}}(\theta \mid x_{(n)}) - \mathcal{N}\left(\hat{\theta}_n, \frac{1}{n} \mathcal{I}^{-1}(\theta_0)\right)}_\mathrm{TV} \rightarrow 0 \quad \text{as } n \rightarrow \infty.
  \end{equation*}
\end{theorem}
The proof is provided in Appendix~\ref{app:proofs} and is accompanied by Lemmas~\ref{lem:bernstein1} and \ref{lem:bernstein2}, also in Appendix~\ref{app:proofs}. 

Theorem~\ref{the:bernstein} requires $\tilde{\Theta}$ to have a non-empty interior in $\mathbb{R}^d$ which covers only certain regimes considered by our methodology. For instance, if $\tilde{\Theta}$ describes a non-negative orthant or a convex cone. However, we may equally study projected posteriors when $\tilde{\Theta}$ has zero measure in $\Theta$; for instance, if $\tilde{\Theta}$ describes the probability simplex. This motivates us to extend these results to Theorem~\ref{the:bernstein_manifold}, below, where we assume $\tilde{\Theta}$ is a $k < d$ dimensional submanifold of $\Theta$. We take motivation from Theorem~1.3 of \cite{winter2023sequential}, who provide a BvM for posterior distributions constrained on a Riemannian manifold. We use the concept of a local chart from differential geometry: for each 
$\theta_0 \in \tilde{\Theta}$, there exists a neighborhood 
$V \subset \mathbb{R}^d$ such that $\varphi : \tilde{\Theta} \cap V \to U \subset \mathbb{R}^k$ is a diffeomorphism onto an open set $U$. This provides local 
Euclidean coordinates for the $k$-dimensional submanifold 
$\tilde{\Theta} \subset \mathbb{R}^d$.
Theorem~\ref{the:bernstein_manifold} therefore establishes posterior convergence in these local coordinates, that is, on $U \subset \mathbb{R}^k$.

\begin{theorem}\label{the:bernstein_manifold}
  Suppose $\tilde{\Theta} \subset \Theta \subset \mathbb{R}^d$ is a $k$-dimensional $\mathcal{C}^2$ embedded submanifold and $\theta_0 \in \tilde{\Theta}^\circ$ is an interior point relative to $\tilde{\Theta}$. 
  Assume that Assumptions~\ref{ass:banach}--\ref{ass:map} and \ref{ass:bernstein} hold for the unconstrained model and that the projection $T_{\tilde{\Theta}}$ is globally $\mathcal{C}^2$ on $\Theta$. Let $(U,\varphi)$ be a local chart on $\tilde{\Theta}$ with $\varphi : U \subset \mathbb{R}^k \to \tilde{\Theta}$, $\varphi(0)=\theta_0$. Denote $u = \varphi^{-1}$ on $\varphi(U)$ and define
  \begin{equation*}
  g : \Theta \to U,
  \qquad
  g(\theta) \coloneq (u \circ T_{\tilde{\Theta}})(\theta).
  \end{equation*}
  For any probability measure $\mu$ on $\Theta$, define its pushforward through $g$ by
  \begin{equation*}
  g_\sharp \mu (B) \coloneq \mu\big(\{\theta \in \Theta : g(\theta)\in B\}\big).
  \end{equation*}
  Let $\tilde{\Pi}_{\tilde{\Theta}}^u(\cdot \mid x_{(n)}) 
  := g_\sharp \Pi_\Theta(\cdot \mid x_{(n)})$
  be the projected posterior in chart coordinates. Then
  \begin{equation*}
  \mathbb{E}_{\mathrm{P}(x_{(n)} \mid \theta_0)}
  \Big\|
  \tilde{\Pi}_{\tilde{\Theta}}^u(\cdot \mid x_{(n)})
  -
  g_\sharp \mathcal{N}\left(\hat{\theta}_n, \frac{1}{n} \mathcal{I}^{-1}(\theta_0)\right)
  \Big\|_{\mathrm{TV}}
  \;\longrightarrow\; 0
  \quad \text{as } n \to \infty.
\end{equation*}
\end{theorem}

For the truncated posterior, an analogous result to Theorem~\ref{the:bernstein} holds when $\theta_0$ lies in the interior of $\tilde{\Theta}$ and $\tilde{\Theta}$ has the same topological dimension as the ambient parameter space $\Theta$. In this setting the constraint is asymptotically inactive as posterior mass concentrates in an $n^{-1/2}$-neighborhood of $\theta_0$ that is contained entirely within $\tilde{\Theta}$. Consequently, the truncated posterior coincides asymptotically with the unconstrained posterior. In contrast, no analogue of Theorem~\ref{the:bernstein_manifold} holds for the truncated posterior when $\tilde{\Theta}$ is an embedded submanifold of lower dimension than $\Theta$. The truncated posterior remains absolutely continuous with respect to the Lebesgue measure on the ambient space, whereas the limiting distribution in Theorem~\ref{the:bernstein_manifold} is supported on the lower-dimensional manifold $\tilde{\Theta}$. As a result, the truncated posterior cannot converge in total variation to the manifold-supported Gaussian limit.

\section{Case Studies} \label{sec:case_studies}

The generality of our method allows application across a wide range of constrained inference problems. We present five case studies across different modeling problems. The first two consider estimation constrained to the probability simplex, with the second imposing additional order constraints. The third provides an example of low-rank covariance estimation. The fourth and fifth focus on two distinct applications of GP regression: GPs subject to inequality and directional constraints. As we have discussed, the projected posterior may inherit many of the computational benefits of the unconstrained posterior; this is particularly so with GP regression where the conditional distribution is analytical. 

\subsection{Case Study 1: Projection on the probability simplex}

We first provide a simple example of the projected posterior in the canonical setting of multinomial probabilities constrained to the simplex. Let
\[
\tilde{\Theta} = \Delta^{d-1} \coloneq \left\{ p \in \mathbb{R}^{d} : p_i \geq 0, \sum_{i=1}^d p_i = 1 \right\},
\]
and suppose that $x_{(n)} \sim \mathrm{Multinomial}(n, p_0)$ for some $p_0 \in \Delta^{d-1}$. Under a Dirichlet prior $\mathrm{Dir}(\alpha)$, the posterior distribution is $\mathrm{Dir}(\alpha + x_{(n)})$. Here, the exact posterior distribution is available in closed form, so the use of projection is not motivated by computational necessity. Rather, we include this example as a transparent illustration of the proposed methodology. Working in a $(d-1)$-dimensional local parametrization, the BvM Theorem implies that, under standard regularity conditions and for $p_0$ in the interior of $\Delta^{d-1}$, the posterior admits the Gaussian approximation
\begin{equation} \label{eqn:bvm_dirichlet_approx}
p \mid x_{(n)} \approx \mathcal{N}\left(p_0,\frac{1}{n}\Sigma(p_0)\right),
\qquad
\Sigma(p_0) = \mathrm{diag}(p_0) - p_0 p_0^\top.
\end{equation}
In practice, $p_0$ is unknown and must be replaced by an estimator. We set $p_0 \approx \mathbb{E}[p \mid x_{(n)}]$ as the posterior mean under the Dirichlet posterior. Note, that we could instead set $p_0 \approx \hat{p}_n = x_{(n)}/n$ as the maximum likelihood estimator. While both choices are asymptotically equivalent, the use of the posterior mean yields a more stable finite-sample approximation and provides a more meaningful comparison with the exact posterior distribution. We treat this Gaussian distribution as an unconstrained proxy supported on $\mathbb{R}^d$, and define the projected posterior as the pushforward of this approximation under a projection map $\mathrm{T}_{\Delta^{d-1}} : \mathbb{R}^d \to \Delta^{d-1}$.

We test the posterior projection under two regimes. In a low-dimensional setting, we take $d=3$ with $p_0 = (0.98,\,0.01,\,0.01)$, with $n = 100$ samples. In a high-dimensional setting, we take $d=50$ with $p_0 = (0.50,\,0.01,\,\ldots,\,0.01)$ and $n = 300$ samples. In both cases, the parameter lies close to a vertex and the covariance structure is highly anisotropic. These parameterizations were chosen to yield $\sim 25\%$ invalid draws when flattened from $\mathbb{R}^d$ onto the $\mathbb{R}^{d-1}$ plane intersecting $\Delta^{d-1}$. We calculate the posterior expected squared error (PESE) and the energy score between the samples and $p_0$ for the true posterior, the BvM-inspired approximation in \eqref{eqn:bvm_dirichlet_approx}, and the projection of this to $\Delta^{d-1}$ under both the Euclidean and Mahalanobis norms. Scores are averaged over 1000 simulations. In both settings, the projected posteriors improve metric scores over the posterior approximation in \eqref{eqn:bvm_dirichlet_approx}. There is very little difference between the Euclidean and Mahalanobis induced projected posteriors. In the low-dimensional regime, this is due to the true parameter lying far in the corner of $\mathrm{T}_{\Delta^{d-1}}$ and so much of the projected mass congregates on the vertex, regardless of the chosen norm. In the high-dimensional regime, the posterior covariance in \eqref{eqn:bvm_dirichlet_approx} is close to identity. In terms of computational speed, both norms have an analytical solution and require very little computation; units are given in nanoseconds.

\begin{table}[t]
\centering
\caption{Comparison of methods across low- and high-dimensional settings. The posterior expected squared error (PESE) is scaled $10^4$, the energy score (Energy) is scaled $10^3$, and the computational time per sample (Comp.) is in units ns.}
\label{tab:case_study}
\begin{tabular}{lcccccccc}
\toprule
& \multicolumn{4}{c}{Low dimension ($d=3$)} 
& \multicolumn{4}{c}{High dimension ($d=50$)} \\
& True & BvM & proj $\norm{\cdot}_2$ & proj $\norm{\cdot}_\Sigma$ 
& True & BvM & proj $\norm{\cdot}_2$ & proj $\norm{\cdot}_\Sigma$ \\
\midrule
PESE 
& 8.42 & 8.49 & 7.78 & 7.78
& 48.7 & 49.1 & 47.6 & 47.8 \\

Energy 
& 12.5 & 12.7 & 12.4 & 12.3
& 34.0 & 34.0 & 33.7 & 33.9 \\

Comp.
& 0.817 & 0.034 & 10.1 & 12.7
& 2.67 & 1.406 & 14.1 & 15.6 \\
\bottomrule
\end{tabular}
\end{table}

\subsection{Case Study 2: Stochastic ordering on a contingency table}

Although the previous case study serves as a simple illustration, the full posterior is easily available and we would not require posterior projections for such a simple problem in practice. However, we may extend this notion of estimation of probabilities on the simplex to estimation of ordered contingency tables. Data in the form of contingency tables arise when individuals are classified by multiple criteria. Often, one or more of the categorical variables have a natural ordering; one example is dose-response studies where the ordered levels correspond to increasing levels of exposure. See \cite{agresti2002analysis} who survey order-constrained statistical methods for contingency tables. 

We consider data from \cite{agresti1998order} on patients with subarachnoid hemorrhage, classified by treatment level and clinical outcome. Let $X \in \{1,\dots,I\}$ denote treatment group (ordered by increasing dose, with $i=1$ placebo) and $Y \in \{1,\dots,J\}$ denote outcome category (ordered from worst to best: death, vegetative state, major disability, minor disability, good recovery). The data are summarized in an $I \times J$ contingency table with counts $n_{ij}$, where $n_{ij}$ is the number of patients in group $i$ with outcome $j$, and $\sum_{i=1}^I \sum_{j=1}^J n_{ij} = n$. Let $n_{i+} = \sum_{j=1}^J n_{ij}$. We assume independent multinomial sampling across treatment groups,
\begin{equation} \label{eqn:so_model}
(n_{i1}, \dots, n_{iJ}) \sim \mathrm{Multinomial}(n_{i+}, \theta_{[i]}), 
\quad i = 1,\dots,I,
\end{equation}
where $\theta_{[i]} = (\theta_{i1}, \dots, \theta_{iJ})$ with $\theta_{ij} = \mathbb{P}(Y=j \mid X=i)$ and $\sum_{j=1}^J \theta_{ij} = 1$. To encode the hypothesis that the relationship between treatment level and patient outcomes is non-decreasing, we impose the stochastic ordering across $i$,
\begin{equation} \label{eqn:so_constraint}
\sum_{k=1}^j \theta_{(i+1)k} 
\;\geq\; 
\sum_{k=1}^j \theta_{ik},
\quad i=1,\dots,I-1,\;\; j=1,\dots,J.
\end{equation}

To implement the posterior projection approach, we first ignore the order constraint \eqref{eqn:so_constraint} to obtain an unconstrained posterior, and then project back into the valid constraint. We assign independent conjugate priors $\theta_{[i]} \sim \mathrm{Dirichlet}(\alpha,\dots,\alpha), \quad i=1,\dots,I,$ with $\alpha=1$. The resulting unconstrained posterior is given as
\begin{equation}
\label{eqn:so_posterior}
\theta_{[i]} 
\sim 
\mathrm{Dirichlet}(n_{i1}+\alpha,\dots,n_{iJ}+\alpha),
\quad i=1,\dots,I.
\end{equation}
We draw independent samples from \eqref{eqn:so_posterior} and project each draw into $\tilde{\Theta} = \{\theta: \eqref{eqn:so_constraint} \text{ holds}\}$ under the Euclidean norm. This choice of norm is motivated by analytical tractability and by Case Study~1, which shows negligible performance differences across norms for probability-simplex problems. As a comparison, we implement the methodology of \cite{laudy2007bayesian}, who enforce the constraint directly by placing a truncated prior on $\theta = (\theta_{[1]}^\top,  \cdots,  \theta_{[I]}^\top )$ using the gamma representation of the Dirichlet distribution restricted to \eqref{eqn:so_constraint}. Posterior inference is then carried out via Gibbs sampling. 

We draw $10^4$ samples from the posterior using both the Gibbs sampler and the posterior projection approach. For the Gibbs sampler, we generate $2 \times 10^4$ samples and discard the first half as burn-in. As is typical in constrained settings, the Gibbs sampler exhibits slow mixing and substantial autocorrelation across components of $\theta$. In contrast, the projection approach yields independent samples by construction, eliminating autocorrelation entirely. Since the cost of projection is negligible, this results in a marked improvement in sampling efficiency. This advantage will become more pronounced as the dimension of the contingency table increases, where mixing for the Gibbs sampler further deteriorates. We compare the two approaches using posterior mean fit and posterior uncertainty. Let $\hat{n}_{ij} = n_{i+} \, \mathbb{E}[\theta_{ij} \mid x_{(n)}]$
denote the posterior expected cell count. The average absolute deviation $(IJ)^{-1} \sum_{i=1}^I \sum_{j=1}^J 
\left| \widehat{n}_{ij} - n_{ij} \right|$ is $16.0$ under the truncated posterior and $5.0$ under the projection approach, indicating substantially improved agreement with the observed table. We also compare posterior uncertainty via the average length of marginal credible intervals for $\theta_{ij}$. The projection approach yields intervals that are, on 
average, $1.25\times$ shorter than those obtained from the Gibbs sampler. We assess frequentist coverage via a simulation study. We generate $10^3$ parameter values for each parameter and simulate a dataset from \eqref{eqn:so_model}. For each dataset, we construct marginal $95\%$ credible intervals for both methods and each $\theta_{ij}$. Empirical coverage probabilities are reported in Table~\ref{tab:so}. The projection-based approach yields coverage close to the nominal level, whereas the Gibbs sampler exhibits systematic deviations from nominal coverage. 

\begin{table}[h]
\centering
\begin{tabular}{|c|ccccc|ccccc|} 
\hline
& \multicolumn{5}{c|}{Constrained Gibbs Sampler} & \multicolumn{5}{c|}{Posterior Projection} \\
\hline
$\theta_{ij}$ & $j=1$ & $j=2$ & $j=3$ & $j=4$ & $j=5$ 
              & $j=1$ & $j=2$ & $j=3$ & $j=4$ & $j=5$ \\
\hline
$i=1$ & $0.513$ & $0.939$ & $0.968$ & $0.938$ & $0.700$
      & $0.954$ & $0.961$ & $0.958$ & $0.948$ & $0.951$ \\
$i=2$ & $0.946$ & $0.974$ & $0.976$ & $0.963$ & $0.948$
      & $0.945$ & $0.957$ & $0.954$ & $0.953$ & $0.954$ \\
$i=3$ & $0.924$ & $0.965$ & $0.969$ & $0.951$ & $0.966$
      & $0.962$ & $0.942$ & $0.943$ & $0.950$ & $0.962$ \\
$i=4$ & $0.598$ & $0.931$ & $0.966$ & $0.888$ & $0.656$
      & $0.946$ & $0.943$ & $0.949$ & $0.952$ & $0.959$ \\
\hline
\end{tabular}
\caption{Frequentist coverage of 95\% credible intervals for components of $\theta$. The Gibbs sampler \citep{laudy2007bayesian} exhibits undercoverage (average $88.4\%$), whereas the posterior projection approach attains near-nominal coverage (average $95.2\%$).}
\label{tab:so}
\end{table}

\subsection{Case Study 3: Low-rank and spiked covariance estimation}

Next, we consider estimation of a low-rank and spiked covariance matrix of the form
\[
\Sigma_\star = U_\star \Lambda_\star U_\star^\top + \sigma^2 I_p,
\qquad
U_\star \in \mathrm{St}(p,r), \quad \Lambda_\star = \mathrm{diag}(\lambda_1,\dots,\lambda_r),
\]
with $\lambda_j \ge 0$ and $r \ll p$. Here, $\mathrm{St}(p,r)$ denotes the Stiefel manifold of orthonormal $r$-frames in $\mathbb{R}^p$, defined as 
\[
\mathrm{St}(p,r)
=
\left\{
\theta \in \mathbb{R}^{p \times r} : \theta^\top \theta = I_r
\right\}.
\]
We define the spiked covariance model class
\[
\mathcal{M}_r
=
\left\{
\Sigma \in \mathbb{S}^p_{++} :
\Sigma = U \Lambda U^\top + \sigma^2 I_p,\;
U \in \mathrm{St}(p,r),\;
\Lambda = \mathrm{diag}(\lambda_1,\dots,\lambda_r),\;
\lambda_j \ge 0
\right\}.
\]
Such models arise naturally in a range of applications where the dominant variability is confined to a low-dimensional subspace, including factor models \citep{fan2013large}, probabilistic principal component analysis \citep{tipping1999probabilistic}, and spatial statistics via fixed-rank kriging \citep{cressie2008fixed}, where large covariance structures are approximated through a relatively small number of basis functions. The decomposition separates a structured signal component $U_\star \Lambda_\star U_\star^\top$ from isotropic noise $\sigma^2 I_p$, and the inferential objective is to recover both the principal subspace and the associated eigenvalues under finite-sample uncertainty. The parameter space underlying this representation is inherently nonconvex; however, projection under the Frobenius norm admits an analytical and unique solution.

Consider the Gaussian sampling model $x_1, \dots, x_n \sim \mathcal{N}_p(0,\Sigma)$,
with the conjugate inverse-Wishart prior $\Sigma \sim \mathcal{IW}_p(\nu_0,\Psi_0)$, yielding the posterior
\[
\Sigma \mid x_{(n)} \sim \mathcal{IW}_p(\nu_0 + n,\ \Psi_0 + S),
\qquad
S = \sum_{i=1}^n x_i x_i^\top.
\]
This posterior is supported on the full space of positive definite matrices and does not encode the low-rank spiked structure of $\Sigma_\star$.
We obtain samples from the projected posterior by mapping each $\Sigma^{(i)}$ onto $\mathcal{M}_r$ via the Frobenius projection
\[
\tilde{\Sigma}^{(i)}
=
\underset{\Sigma \in \mathcal{M}_r}{\arg\min}
\left\|
\Sigma^{(i)} - \Sigma
\right\|_F^2,
\]
with analytical solution provided in Proposition~\ref{prop:frob_spiked_projection}.

\begin{proposition} \label{prop:frob_spiked_projection}
Let $\Sigma^{(i)} \in \mathbb{R}^{p \times p}$ be symmetric with spectral decomposition
\[
\Sigma^{(i)} = U^{(i)} \,\mathrm{diag}(\alpha_1^{(i)},\dots,\alpha_p^{(i)})\, {U^{(i)}}^\top,
\qquad
\alpha_1^{(i)} \ge \cdots \ge \alpha_p^{(i)},
\]
and write $U_r^{(i)} = (u_1^{(i)},\dots,u_r^{(i)})$ for the matrix of its leading $r$ eigenvectors. Then a Frobenius projection of $\Sigma^{(i)}$ onto $\mathcal{M}_r$ is given by $\tilde{\Sigma}^{(i)}
=
U_r^{(i)} \,\mathrm{diag}(\hat\lambda_1^{(i)},\dots,\hat\lambda_r^{(i)})\, {U_r^{(i)}}^\top
+
\hat\sigma^{2(i)} I_p$, where
\[
\hat\sigma^{2(i)}
=
\frac{1}{p-r}\sum_{j=r+1}^p \alpha_j^{(i)},
\qquad
\hat\lambda_j^{(i)}
=
\max\left(\alpha_j^{(i)}-\hat\sigma^{2(i)},\,0\right),
\quad j=1,\dots,r.
\]
\end{proposition}

Proof is provided in Appendix~\ref{app:proofs}.

\begin{figure}[h]
    \centering
    \includegraphics[width=\linewidth]{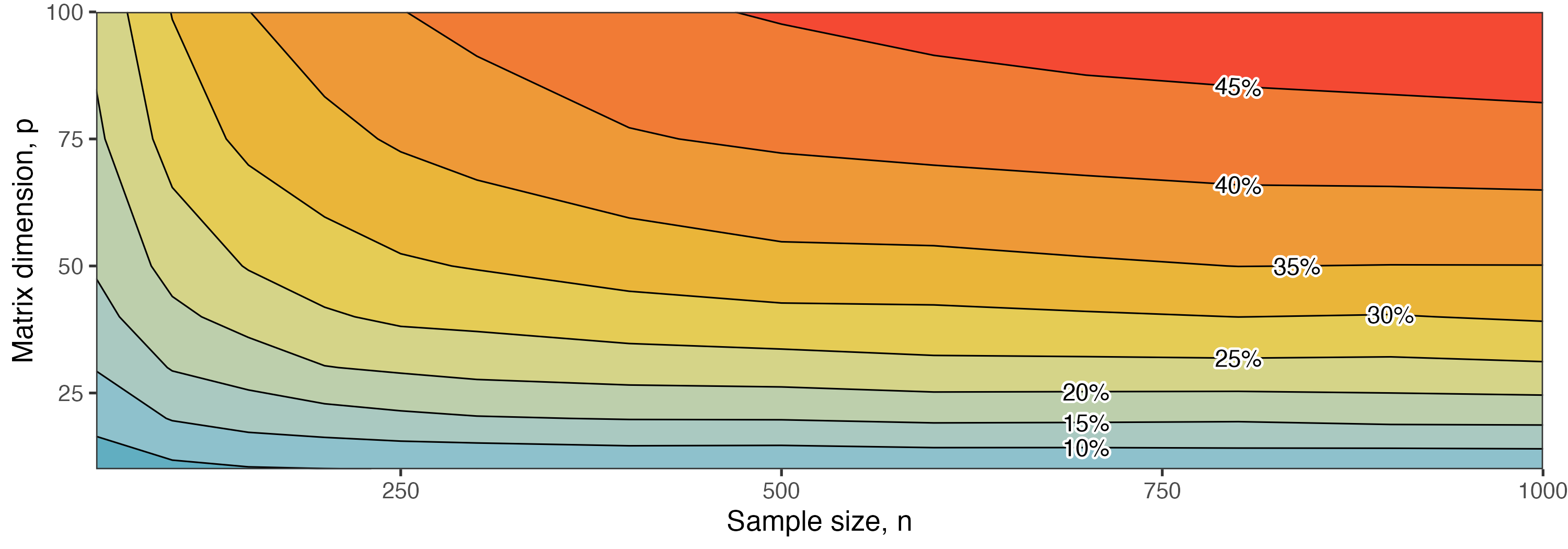}
    \caption{Average percentage reduction in posterior expected Frobenius loss achieved by the projected posterior relative to the unconstrained posterior in the rank-$3$ spiked covariance setting, over a grid of sample sizes $n$ and matrix dimensions $p$. Results are based on $100$ Monte Carlo replicates at each grid point. Larger values indicate greater improvement from projection.}
    \label{fig:lowrank_frobenius_contour}
\end{figure}

We consider a true parameterization given by a fixed rank $r=3$, with eigenvalues $\{\lambda_1, \lambda_2, \lambda_3\} = (6, 3, 1.5)$ and noise variance $\sigma^2 = 1$. Define a grid of sample sizes $50 \leq n \leq 1000$ and ambient dimension $10 \leq p \leq 100$. For each pair $(n, p)$ we generate data $x_{i} \in \mathbb{R}^{n \times p}$ from the corresponding Gaussian model with true covariance $\Sigma_0$ with $U_\star$ sampled uniformly from $\mathrm{St}(p,r)$. Given the data, we construct the unconstrained posterior for $\Sigma$ under an inverse-Wishart prior specified by $\nu_0 = p+2$ and $\Psi_0 = I_p$, and obtain a projected posterior by mapping draws onto $\mathcal{M}_r$. At each value of $(n, p)$, we compute the posterior expected Frobenius loss $\mathbb{E}\!\left[ \|\Sigma - \Sigma_0\|_F \mid x_{(n)} \right]$ 
for both the unconstrained and projected posterior distributions, approximated via Monte Carlo samples. Figure~\ref{fig:lowrank_frobenius_contour} reports the loss percentage reduction achieved by projection, the values of which are averages over $100$ Monte Carlo replicates. Figure~\ref{fig:lowrank_frobenius_contour} shows that the projected posterior yields increasingly substantial gains as both the sample size $n$ and ambient dimension $p$ grow. The broad pattern is monotone in the expected direction: when the problem is more strongly high-dimensional, the structural gain from enforcing the low-rank constraint is more pronounced, and this improvement becomes clearer as the sample size increases. There is also a visible square-root sample-size effect, with the contour geometry exhibiting the familiar $n^{1/2}$ scaling expected from posterior concentration. In practical terms, this indicates that the benefit of projection is modest in smaller or lower-dimensional settings, but becomes materially larger once the signal from posterior contraction is sufficient for the low-rank structure to be exploited effectively.

\subsection{Case Study 4: Linear inequality constrained GPs} \label{sec:inequality}

\begin{figure}[b!]
  \centering
  \includegraphics[width=\textwidth]{"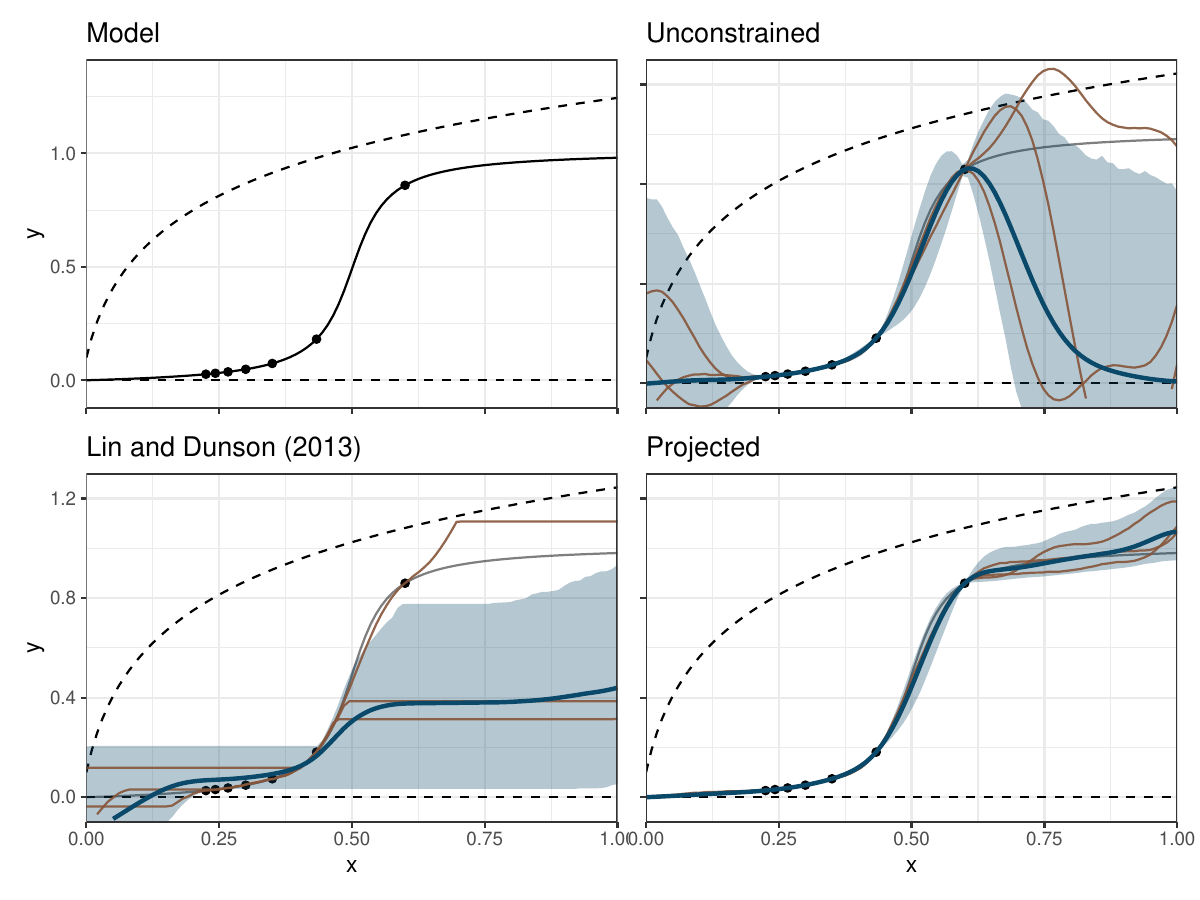"}
  \caption{The model of \cite{agrell2019gaussian} and three different functional estimations, posterior expectations are blue solid lines, and bands denote 90\% predictive intervals. Top left: the solid line is the true function, the dashed lines the bounds, and the points the observed data. Top right: the fit of a zero-mean GP with no monotonicity or boundedness constraints, parameterized by a squared-exponential kernel with parameters $\sigma = 0.5$ and $l = 0.1$. Bottom left: monotonic projections of the unconstrained GP (top right) as per \cite{lin2014bayesian}. Bottom right: monotonic and bounded projections of the unconstrained GP (top right) in the Banach space defined by $\norm{\cdot}_\Sigma$ in \eqref{eqn:mahalnobis}, with $\Sigma$ set as the conditional variance of the unconstrained GP.}
  \label{fig:monotone}
\end{figure}

In many nonparametric regression problems, it is desirable to constrain the function to satisfy shape or inequality constraints. We now demonstrate our methodology in the case study used in \cite{agrell2019gaussian}, which involves an unknown function with a linear inequality constraint. We use the projected posterior under a GP prior. Linear inequality constraints imply that $\tilde{\Theta}$ is convex \citep{boyd2004convex}, thus Assumption~\ref{ass:map} is satisfied when $\Theta$ is a Hilbert space endowed with an inner product, via the Hilbert Projection Theorem. \cite{agrell2019gaussian} define the true function over $x$ as $f(x) = \frac{1}{3}[\tan^{-1}(20x - 10) - \tan^{-1}(-10)]$ for $x \in [0,1]$, and impose a minimum bound of $l(x) = 0$, and an upper bound of $u(x) = \frac{1}{3}\log(30x + 1) + 0.1$. Seven noiseless observations are provided from $f(x)$ at the locations $x_i = 0.1 + 1/(i+1)$ for $i \in [1, \dots, 7]$. The true model $f(x)$, the lower and upper bounds, and the observations are shown in the top left panel of Figure~\ref{fig:monotone}.

In \cite{agrell2019gaussian}, the GP fit is constrained to satisfy monotonicity $\mathrm{d}f/\mathrm{d}x \geq 0$ and boundedness $l(x) \leq f(x) \leq u(x)$. Furthermore, the unconstrained GP is parameterized as a zero-mean process with covariance given by a squared-exponential kernel $k_{\mathrm{SE}}(x, x') = \sigma^2 \exp(-\frac{1}{2}(\frac{x - x'}{l})^2)$ with amplitude $\sigma = 0.5$ and length scale $l = 0.1$. The fit of the unconstrained GP model is shown in the top right panel of Figure~\ref{fig:monotone}; the thick blue line is the posterior mean, the shaded region represents the 90\% predictive interval, and the yellow lines are three samples drawn randomly. In plotting this figure, we drew $1000$ independent samples; none respect the monotonicity and boundedness desired by the model setup, so rejection sampling (i.e. sampling from the truncated posterior) is highly inefficient. In all plots in Figure~\ref{fig:monotone} we approximate the infinite-dimensional GP with 100 points spaced evenly over the domain.

\cite{lin2014bayesian} present a methodology to define a monotonically projected GP by projecting samples $w(x) \sim \mathcal{GP}(\mu(x), k(x, x'))$, indexed on $\mathcal{X} \in \mathbb{R}^d$, from the unconstrained GP into the monotonic cone $\mathcal{M} = \{x_i \leq x_{i+1}\}$. Focusing here on the case of $d = 1$, they define their projection operator by
\begin{equation} \label{eqn:lin_dunson}
  \mathcal{T}_\mathcal{M}(w(x)) = \underset{F \in \mathcal{M}}{\arg\min} \int_{\mathcal{X}} (w(x) - F(x))^2 \; \mathrm{d}x = \underset{v \geq x}{\inf} \; \underset{u \leq x}{\sup}\; \frac{1}{v-u} \int^v_u w(x) \; \mathrm{d}x.
\end{equation}
The posterior moments of the \cite{lin2014bayesian} model are shown in the bottom left panel of Figure~\ref{fig:monotone}, and the samples shown correspond to the projection of the samples shown in the unconstrained model. Sample projection is an embarrassingly parallel problem and, here, requires $\sim 5 \; \mathrm{ms}$ per sample on an Apple M2 Pro processor with 32GB RAM. Although \cite{lin2014bayesian} provides a neat analytical solution to monotonic regression, the projected posterior is unfaithful to the initial prior specifications, as seen by the discontinuous derivatives in the samples. The model of \cite{lin2014bayesian} is subsumed by our theory by setting $\tilde{\Theta} = \mathcal{M}$ and specifying the Banach space by the Euclidean norm $\norm{\cdot}_2$, and so the projection does not respect correlation between locations in $\mathcal{X}$. Furthermore, by requiring the posterior projections to respect the bounds, \eqref{eqn:lin_dunson} no longer holds as an analytical solution, and so this theory does not satisfy the full requirements of the problem.

We project the unconstrained GP using the projection \eqref{eqn:proj_operator} depending on $\norm{\cdot}_\Sigma$ as defined in \eqref{eqn:mahalnobis} in Section~\ref{sec:norm}, where here $\Sigma$ is the conditional variance of the unconstrained GP. The results are shown in the bottom right panel of Figure~\ref{fig:monotone}. Due to the convexity of $\tilde{\Theta}$ in this problem, we may guarantee both the existence and uniqueness of the solution. It is apparent in Figure~\ref{fig:monotone} that we successfully obey the monotone and boundedness constraints. Further, each of our generated samples (also projected from the unconstrained samples) provides a much closer representation to the prior beliefs imparted by the squared-exponential, thus providing evidence of the value of appropriate norm specification. Projections were calculated by \texttt{cvxr}, a bespoke package for convex optimization \citep{fu2017cvxr}; average computation of the unconstrained sample required $\sim0.7 \; \mathrm{ms}$ and projection of each sample required $\sim 40 \; \mathrm{ms}$. With the aid of \texttt{cvxr}, the extra code required for the projections is simple and only comprises $\sim 4$ extra lines of code and readily integrates with standard GP workflows; our code is available at \texttt{github/astfalckl/projector}, and CVX also has packages in \texttt{MATLAB} and \texttt{python} that efficiently solve convex optimization problems \citep[see][]{cvx,diamond2016cvxpy}. This stands in comparison to methodologies such as \cite{agrell2019gaussian} or \cite{wang2016estimating}, bespoke solutions that require more specialized and problem-specific implementation.

\subsection{Case Study 5: Computer emulation of directional outputs} \label{sec:heading}

Finally, we present an application of our methodology to the emulation of a computer model with directional outputs. Directional quantities are not straightforward to represent in a statistical model for a number of reasons. First, as the direction $\theta$ `wraps' around $0$ and $2\pi$, the distance on the circle must be parameterized appropriately, such as geodesic distance, and covariance functions must be defined that are positive semi-definite over the circular domain \citep{gneiting2013strictly}. Although this solves the problem when the inputs are directional \citep[see][]{astfalck2019emulation}, for circular outputs, it remains difficult to define a coherent statistical process directly. The most common methods wrap a distribution defined in $\mathbb{R}$ around the circle \citep{jona2012spatial}, or similar to what we propose, project a distribution from $\mathbb{R}^2$ onto the circle \citep{wang2014modeling,mastrantonio2016spatio}. In fact, as with the methodology of \cite{lin2014bayesian} in Section~\ref{sec:inequality}, the methodology of \cite{wang2014modeling} can be seen as a special case of ours with the norm of the Banach space defined by $\norm{\cdot}_2$.

Directional quantities are a special case of a Stiefel manifold. When $p=1$, the manifold $\mathrm{St}(1, m)$ describes a $(m-1)$-dimensional hypersphere, for example, $m=2$ defines a single direction in $\mathcal{S} \subset \mathbb{R}^2$ as with our example of vessel heading, and $m=3$ describes a bearing in $\mathcal{S}^2 \subset \mathbb{R}^3$ as would be experienced in flight navigation. Proposition 4.9 of \cite{eleonora2017probability} states that the projection, with respect to the trace inner product, of any $\theta \in \Theta_\mathrm{p} = \{\theta \in \mathbb{R}^{m \times p} : \mathrm{rank}(\theta) = p\}$ into $\mathrm{St}(p, m)$ is unique. When $p=1$, the trace inner product induces the norm $\norm{\cdot}_2$ and $\Theta_\mathrm{1} = \mathbb{R}^m \; \backslash \; \{0\}$; that is, the only point that does not respect uniqueness is the origin. We generalize the conditions under which the projection is unique for the norm $\norm{\cdot}_\Sigma$ for $p=1$.Projection under $\norm{\cdot}_\Sigma$ onto the sphere $\mathrm{St}(1,m) = \mathcal S^{m-1}$ is equivalent to Euclidean projection onto an ellipsoid obtained by linearly transforming the sphere by $\Sigma^{-1/2}$. Non-uniqueness occurs only when $\theta$ lies entirely in the subspace orthogonal to the major axis of the ellipsoid (i.e., the minor axes) and bounded inside the corresponding elliptical cross-section. Formally, define $\varphi_1, \cdots, \varphi_{m-1}$ and $\lambda_1, \cdots, \lambda_{m-1}$ as the first $m-1$ eigenvectors and eigenvalues of the matrix $\Sigma^{-1}$ that defines $\norm{\cdot}_\Sigma$, and define 
\begin{equation} \label{eqn:ellipse_set}
  \Theta_{\varphi} = \left\{\theta \in \mathbb{R}^m: \theta = \sum_{i = 1}^{m-1} t_i \varphi_i, \sum_{i=1}^{m-1} t_i^2 \lambda_i < 1 \right\}.
\end{equation}
For example, when $m=2$ as in our application, \eqref{eqn:ellipse_set} 
corresponds to the locations bounded within the circle in the linear subspace that runs perpendicular to the major axis of $\Sigma$. Similarly, when $m=3$, \eqref{eqn:ellipse_set} describes the locations bounded within the sphere and on the plane perpendicular to the major axis of $\Sigma$. The projection into $\mathrm{St}(1, m)$ is unique for all $\theta \in \Theta = \mathbb{R}^m \; \backslash \; \Theta_{\varphi}$. To satisfy Assumption~\ref{ass:map}, the unconstrained posterior must place its entire measure on $\mathbb{R}^m \; \backslash \; \Theta_{\varphi}$. In this application, we wish to build a GP emulator to the computer model, and so the unconstrained posterior is continuous with respect to the Lebesgue measure $\mu$ on $\mathbb{R}^m$. As $\mu(\Theta_{\varphi}) = 0$, sampling from $\mathbb{R}^m$ is equivalent, with respect to $\mu$, to sampling from $\mathbb{R}^m \; \backslash \; \Theta_{\varphi}$.

\begin{figure}[b!]
  \centering
  \includegraphics[width=\textwidth]{"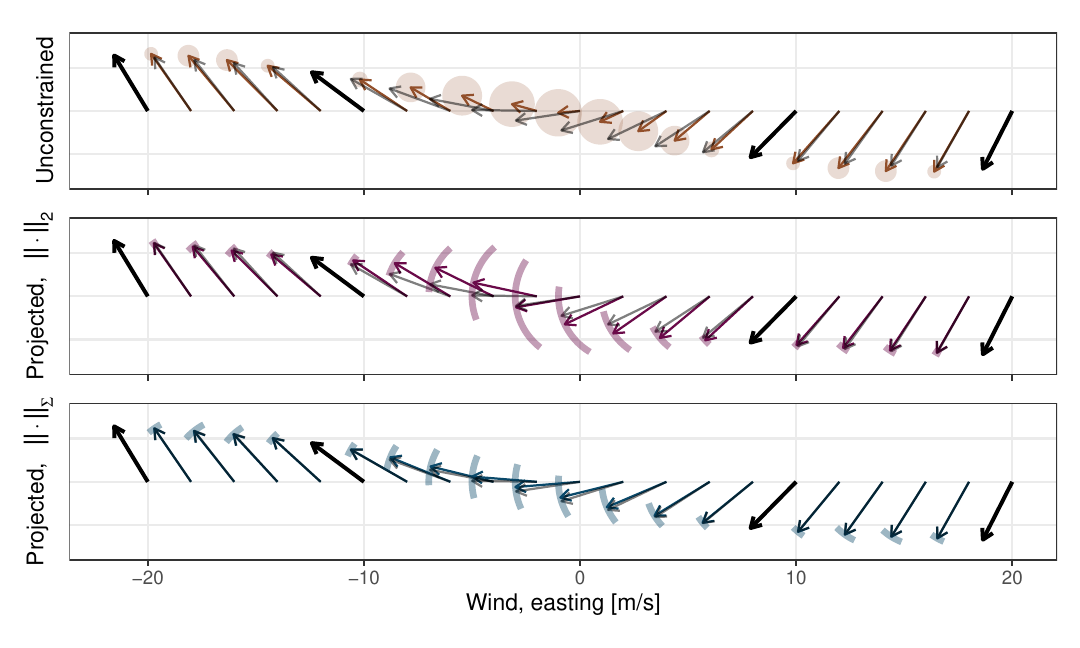"}
  \caption{GP emulation of a computer simulation of vessel headings over a single input with a bivariate GP and the corresponding projections onto the unit circle, projected with respect to $\norm{\cdot}_2$ (middle) and $\norm{\cdot}_\Sigma$ (bottom). Observed values from the computer simulation are shown by the thick black arrows, the respective mean predictions are shown by the colored arrows, the true (unobserved) values from the computer simulation are shown by the thin black arrows. Prediction uncertainty is denoted by the shaded regions: shaded circles for the bivariate GP, and shaded arcs for the projected processes.}
  \label{fig:wind_1d}
\end{figure}

The computer simulator of interest models a moored vessel that is free to weathervane, or rotate, about its mooring. The vessel is subject to wind and current forcings, and achieves a heading that balances the rotational moments on the vessel, induced by wind and current forcing measured as vectors with components in the easting and northing directions. See \cite{milne2016validation} for further details and an example of the application. We show a simple example of the output of the model with respect to a single input in Figure~\ref{fig:wind_1d}. The thick black arrows are the model output at wind easting components of $-20, -10, 10 $ and $ 20 \; \mathrm{ms}^{-1}$, with a constant wind northing component of $10 \; \mathrm{ms}^{-1}$ and current easting and northings of $0.3 \; \mathrm{ms}^{-1}$. The model output is a direction and the arrows lie on the unit circle, although the choice of arrow length is arbitrary. In the top plot of Figure~\ref{fig:wind_1d}, we model the directional output as a bivariate GP, assumed independent between the output dimensions, and predict the computer model over a dense grid of wind easting values. The true directions, unobserved by the GP, of the model are shown by the thin black arrow, the predicted values are the colored arrows, and the uncertainty of the prediction is represented by the shaded regions that denote the $80\%$ centered prediction interval. As expected, the GP predictions do not lie in the unit circle. \cite{wang2014modeling} propose a methodology that projects the samples from the bivariate GP onto the unit circle, in effect, sampling the angles of the bivariate vectors. As noted above, this is equivalent to our methodology in $\norm{\cdot}_2$, and is shown in the middle plot of Figure~\ref{fig:wind_1d}. The bottom plot of Figure~\ref{fig:wind_1d} shows the result of projecting samples from the bivariate GP with respect to $\norm{\cdot}_\Sigma$. Interpretation of the middle and bottom plots is much the same as the top; although uncertainty of the directional quantities is now represented by the shaded arcs around the predictions. Our methodology supports the use of the norms $\norm{\cdot}_2$ and $\norm{\cdot}_\Sigma$; in this instance, a better mean performance is observed under $\norm{\cdot}_\Sigma$ with a root-mean-square error of $1.9 \; \mathrm{deg}$, as opposed to under $\norm{\cdot}_2$ with root-mean-square error of $6.5 \; \mathrm{deg}$. The differences are most noticeable between the values of $-10$ and $10\;\mathrm{ms}^{-1}$. The base GP predictions required $\sim0.40 \; \mathrm{ms}$ per sample and the projections required $\sim7.6 \; \mathrm{ms}$.

\begin{figure}[t!]
  \centering
  \includegraphics[width=\textwidth]{"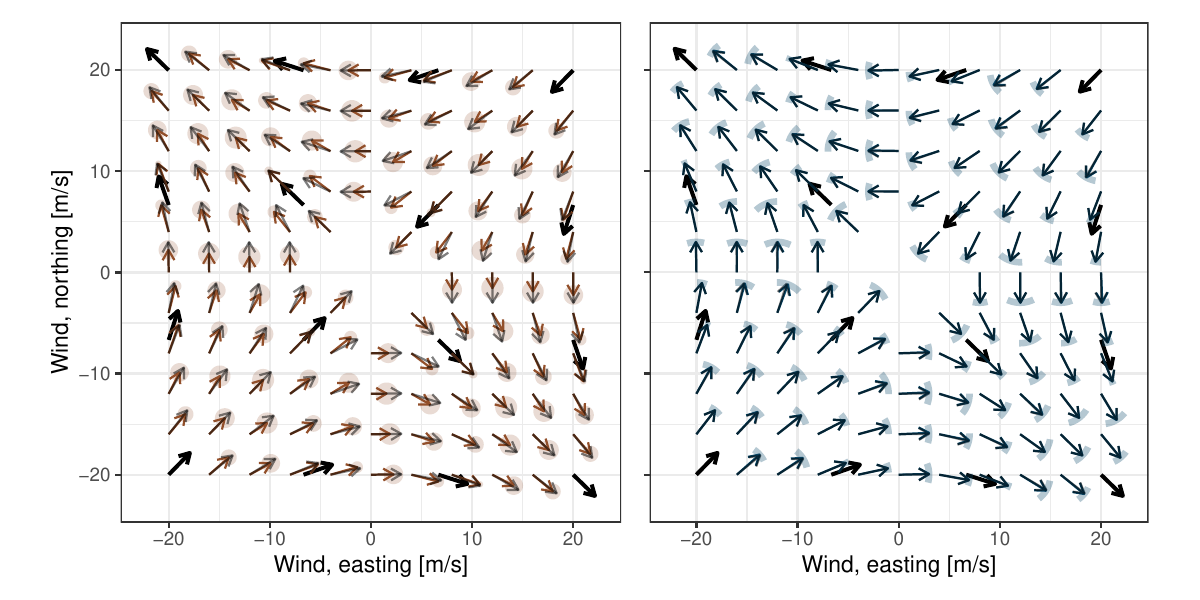"}
  \caption{A 2D slice in the input space of a computer emulation of a vessel heading model, emulated over a four-dimensional input space represented by the wind and current vector forcings. The slice corresponds to zero current forcing. The left plot is the emulation with a simple bivariate GP, and the right the projected process with respect to $\norm{\cdot}_\Sigma$. Interpretation of the plot is as in Figure~\ref{fig:wind_1d}.}
  \label{fig:wind_2d}
\end{figure}

We now emulate the full computer model over the four dimensional input space defined by the wind and current vector components. We train on simulation outputs from a regular grid between $-20$ and $20 \; \mathrm{ms}^{-1}$ for the wind components and $-0.5$ and $0.5 \; \mathrm{ms}^{-1}$ for the current components. In regions of zero forcing, or when the wind and current fields directly oppose each other, the model is ill-defined. Consequently, we exclude these regions from the predicted locations. Figure~\ref{fig:wind_2d} shows a slice of the emulated output corresponding to the wind-only loading case, that is, zero current components. The interpretation of this graph is the same as in Figure~\ref{fig:wind_2d}, with the unconstrained bivariate GP shown on the left and the projected process on the right. For this particular slice of the input space, the origin corresponds to the zero-loading case, and so it, and surrounding regions, is not included in the predicted locations. As in the example provided in Figure~\ref{fig:wind_1d}, the bivariate GP predictions do not lie on the unit circle, although there is valuable information in the predicted angles. In the plot of the projected process, the true simulator values have also been plotted but are almost completely covered by the predicted expectations. The base GP predictions required $\sim0.96 \; \mathrm{ms}$ per sample and the projections required $\sim300 \; \mathrm{ms}$. As our methodology subsumes that of \cite{wang2014modeling}, it inherits the ability to model directional bimodality and skewness through richer classes of unconstrained GPs that are not isotropic in the output dimensions, allowing the representation of more complex facets of the directional process.

\section{Relaxations and Extensions} \label{sec:relax}

The preceding development imposes conditions that ensure the projected posterior is well defined as a deterministic push-forward of the unconstrained posterior under a single-valued metric projection onto a closed constraint set. These assumptions provide a canonical construction and straightforward computation, but they are not strictly necessary for a meaningful constrained update. This section examines extensions in which either uniqueness or existence of the projection may fail. In each case, the projected posterior can still be defined in a measure-theoretic sense, though additional modeling or algorithmic choices may be required.

\subsection{Non-unique Projections} \label{sec:non_unique}

Assumption~\ref{ass:map} ensures that the projection operator $T_{\tilde{\Theta}}(\theta)$ is single-valued. In this case $T_{\tilde{\Theta}}$ is an ordinary measurable map and the projected posterior is unambiguously defined as the push-forward measure in \eqref{eqn:proj_operator}. Here, the projected posterior is entirely determined by the geometry of $(\Theta,\|\cdot\|)$ and the constraint set $\tilde{\Theta}$. From a computational perspective, samples from the unconstrained posterior $\{\theta^{(m)}\}$ can be projected pointwise, yielding $\{T_{\tilde{\Theta}}(\theta^{(m)})\}$, without any ambiguity or dependence on implementation details. The construction is therefore deterministic, reproducible and computational difficulty simply depends on the geometry of the projection. To relax Assumption~\ref{ass:map} implies that the minimizer need not be unique and the projection becomes set-valued,
\[
S_{\tilde{\Theta}}(\theta)
=
\Bigl\{
\tilde{\theta}\in\tilde{\Theta} :
\|\theta-\tilde{\theta}\|
=
\mathrm{dist}(\theta,\tilde{\Theta})
\Bigr\}.
\]
This does not invalidate the theoretical construction of a projected posterior; however, a deterministic push-forward is no longer available and so practical implementation is more nuanced. Instead, one may introduce a Markov kernel 
$K:\Theta\times\mathcal{B}_{\tilde{\Theta}}\to[0,1]$ 
such that for each $\theta\in\Theta$, the probability measure $K(\theta,\cdot)$ is supported on $S_{\tilde{\Theta}}(\theta)$. The projected posterior is then defined as
\begin{equation} \label{eqn:markov}
  \tilde{\Pi}_{\tilde{\Theta}}(\tilde{B}\mid x_{(n)})
=
\int_\Theta
K(\theta,\tilde{B})\,
\mathrm{d}\Pi_\Theta(\theta\mid x_{(n)}),
\end{equation}
generalizing the definition in \eqref{eqn:proj_posterior}. The substantive issue is not existence but uniqueness. When the projection is set-valued, any implementation must resolve multiplicity either explicitly or implicitly. Explicit specification of some $K(\theta,\cdot)$ is likely too arduous for the modeling problems for which projected posteriors are appreciably beneficial (as compared to classic solutions). Implicit specification imposes an optimization routine that implicitly defines some $K(\theta,\cdot)$. For instance, deterministic optimization routines may induce a kernel as a function of algorithmic details such as initialization or numerical tolerances; stochastic routines introduce additional randomness. In either case, the solver implicitly defines a selection mechanism and hence a Markov kernel. This may introduce implementation dependent behavior that is not intrinsic to the model; however, in many regular settings such dependence may be negligible and represent an acceptable trade-off for the afforded computational convenience. 

\subsection{Open Constraint Sets}

Assumption~\ref{ass:subset} requires $\tilde{\Theta}$ to be closed. When $\tilde{\Theta}$ is open, the metric projection need not exist, since the distance $\mathrm{dist}(\theta,\tilde{\Theta}) = \inf_{\tilde\theta\in\tilde{\Theta}} \|\theta-\tilde\theta\|$ may only be attained on the boundary $\partial\tilde{\Theta}$. In contrast to Section~\ref{sec:non_unique}, the difficulty is therefore one of existence rather than uniqueness. There are several principled, practical extensions available. The appropriate choice depends on whether the open constraint reflects a genuine topological restriction or a practical modeling preference for interior solutions.

The most minimal modification is to replace $\tilde{\Theta}$ by its closure $\tilde{\Theta}^\mathrm{cl}$ and define
\[
T_{\tilde{\Theta}^\mathrm{cl}}(\theta)
=
\arg\min_{\tilde\theta\in\tilde{\Theta}^\mathrm{cl}}
\|\theta-\tilde\theta\|.
\]
The existence of a minimiser is restored under the same conditions as in Assumption~\ref{ass:subset}. The main implication is that the resulting posterior may place mass on $\partial\tilde{\Theta}$. In many regular parametric settings, if the unconstrained posterior is absolutely continuous, this boundary mass is asymptotically negligible. However, if the true parameter lies on or near the boundary, or in small sample regimes, the behavior may differ materially from a strictly interior constraint and be unsuitable.

The next modification is to explicitly enforce a strict interior condition via a boundary modification. The simplest option is via $\varepsilon$-thickening where we project into the closed set $\tilde{\Theta}_\varepsilon = \{\theta : \mathrm{dist}(\theta,\tilde{\Theta}) \le \varepsilon\}$, for some user-specified $\varepsilon$. Should the existence of such tolerance be natural, e.g. a minimum tolerated bound on an intensity or probability, this is a simple modification. Such modifications are common in practice; for example, small positive offsets are routinely introduced when applying log-transformations to zero counts, effectively replacing a hard boundary at zero by a numerically stable interior approximation. Alternatively, one may project onto $\tilde{\Theta}^{\mathrm{cl}}$ and then apply a Markov kernel, similar to \eqref{eqn:markov}, that redistributes any boundary mass into the interior, for example by a small inward probabilistic perturbation. This can result in a softer boundary than $\varepsilon$-thickening, although requires more detailed specification.

Finally, a more structural alternative replaces hard projection onto $\tilde{\Theta}$ by penalization. One may define, for example, the penalized projection
\[
T_{\tilde{\Theta}^\mathrm{pen}}(\theta)
=
\arg\min_{\tilde{\theta} \in \tilde{\Theta}}
\Bigl\{
\|\theta-\tilde{\theta}\| + \lambda \phi(\tilde{\theta})
\Bigr\},
\]
where $\phi(\tilde{\theta})$ is zero on $\tilde{\Theta}$ and 
$\phi(\tilde{\theta}) \to \infty$ as $\mathrm{dist}(\tilde{\theta}, \partial \tilde{\Theta}) \to 0$ for $\tilde{\theta} \in \tilde{\Theta}$. Such behavior is achieved, for instance, with log-barrier or reciprocal-distance penalties. This yields a deterministic interior solution and avoids explicit projection onto a set. The approach is smooth and often computationally stable, but it modifies the geometry of the problem and is no longer a projection onto $\tilde{\Theta}$ via \eqref{eqn:proj_operator}.

\section{Discussion} \label{sec:discussion}

This paper presents a unified framework for Bayesian inference in constrained parameter spaces via posterior projection. Unlike existing approaches, which often rely on bespoke methodologies tailored to specific constraints, our method generalizes to a wide range of parameter constraints, is robust to the choice of norm, and is easily implemented, thus enabling practical modeling in spaces where existing priors or constraints would otherwise be infeasible. We support the methodological contribution with theoretical guarantees, including posterior consistency, contraction, and optimal asymptotic coverage via a BvM Theorem. In addition, we demonstrate the practicality and flexibility of our approach through case studies. The proposed methodology avoids the challenges of truncation-based methods, which can be computationally prohibitive to sample from or infeasible for measure-zero constraints, and has significant implications for both theory and practice. Practically, the flexibility to define projections over arbitrary norms allows practitioners to tailor the methodology to specific problem domains. By avoiding rejection sampling or bespoke MCMC schemes, the projected posterior offers a computationally efficient alternative that scales well to higher dimensions. We note that this methodology does not protect against model misspecification. Imposing a parameter constraint represents a strong prior belief on the model, and the burden of correctly specifying constraints lies with the modeler not the model.

We provide some asymptotic guarantees and empirical evidence, but there remain exciting avenues for further exploration. From a theoretical perspective, studying $\theta \in \partial \tilde{\Theta}$, that is, when $\theta$ lies on the boundary of $\tilde{\Theta}$ is intriguing. For instance, if $\Pi_\Theta(\theta \mid x_{(n)})$ has a density that is positive and continuous in a neighborhood of $\theta_0$. Under standard local regularity and identifiability conditions on the likelihood, the truncated posterior still contracts to $\theta_0$, even when $\theta_0\in\partial\Theta$, at the usual contraction rate. However, the truncated posterior remains absolutely continuous on $\Theta$ and hence assigns zero mass to $\partial\Theta$; consequently, when $\theta_0\in\partial\Theta$ it concentrates \emph{near} the boundary but never \emph{on} it. By contrast, the projected posterior can induce a singular component supported on $\partial\Theta$, which can materially affect finite-$n$ summaries (posterior mean/median, credible sets), even though both approaches contract to the same boundary truth asymptotically. Additionally, we conjecture that stronger results on frequentist coverage of the credible intervals can be shown. Specifically, let $q_\alpha(x_{(n)})$ and $\tilde{q}_\alpha(\mathrm{x}_{(n)})$ denote the $\alpha$th quantile of the unconstrained and projected posterior for $\alpha \in (0,1)$, respectively, and define the coverage probabilities $c_\alpha = \mathrm{P}(q_{\alpha/2}(x_{(n)}) < \theta_0 < q_{1 - \alpha/2}(x_{(n)}) \mid \theta_0)$ and $\tilde{c}_\alpha = \mathrm{P}(\tilde{q}_{\alpha/2}(x_{(n)}) < \theta_0 < \tilde{q}_{1 - \alpha/2}(x_{(n)}) \mid \theta_0)$. We hypothesize that $|c_\alpha - \tilde{c}_\alpha| \rightarrow 0$ as $n \rightarrow \infty$, providing further justification for the validity of projected posteriors in large-sample settings. From an application perspective, there is potential to expand the framework to more complex constraint structures, such as those arising in hierarchical models where constraints operate at multiple levels, for example, in the application of \cite{astfalck2024coexchangeable}. Identifying broader families of constraints that satisfy Assumption~\ref{ass:map} or exploring constraints with dynamic or time-varying characteristics could also enhance the versatility of the approach.

In summary, posterior projection offers a rigorous, flexible, and computationally efficient framework for inference under constraints. Its combination of generality, theoretical guarantees, and practical ease of implementation makes it a valuable addition to the Bayesian toolkit, with wide relevance to modern statistical and machine learning applications.

\section*{Acknowledgement}
LA and EC are supported by the ARC ITRH for Transforming energy Infrastructure through Digital Engineering (TIDE), Grant No. IH200100009. DS acknowledges support from the Statistical and Applied Mathematics Institute (SAMSI), Grant No. DMS-1638521. DD was partially supported by the National Institutes of Health (grant ID R01ES035625), by the European Research Council under the European Union’s Horizon 2020 research and innovation program (grant agreement No 856506), and by the Office of Naval Research (N00014-21-1-2510).

\bibliographystyle{biometrika}
\bibliography{references}

\appendix

\section{Proofs} \label{app:proofs}

\begin{proof} \textbf{(of Theorem~\ref{the:decision})}
  For $\theta \in \Theta$, define the loss function
  \begin{equation*}
    \tilde{\mathrm{L}}(\theta, \xi) = \mathrm{L}(\mathrm{T}_{\tilde{\Theta}}(\theta), \xi) + h(\theta, \mathrm{T}_{\tilde{\Theta}}(\theta))
  \end{equation*}
  where $h(\cdot, \cdot)$ is a non-negative function with $h(\theta, \theta) = 0$. A sufficient choice is $h(\cdot, \cdot) = \mathrm{dist}(\cdot, \cdot)$ where $\mathrm{dist}(\cdot, \cdot)$ is defined as in \eqref{eqn:proj_operator}. When $\theta \in \tilde{\Theta}$, we have $\tilde{\mathrm{L}}(\theta, \xi) = \mathrm{L}(\theta, \xi)$ and so
  \begin{equation} \label{eqn:bayes_loss_projected}
    \int_\Theta \tilde{\mathrm{L}}(\theta, \xi) \; \mathrm{d}\tilde{\Pi}_{\tilde{\Theta}}(\theta \mid x_{(n)}) = \int_\Theta \mathrm{L}(\theta, \xi) \; \mathrm{d}\tilde{\Pi}_{\tilde{\Theta}}(\theta \mid x_{(n)})
  \end{equation}
  as $\tilde{\Pi}_{\tilde{\Theta}}(\theta \mid x_{(n)})$ is only measurable in $\tilde{\Theta}$. The expected loss of $\tilde{\mathrm{L}}(\theta, \xi)$ under the unconstrained posterior is
  \begin{equation} \label{eqn:bayes_loss_unconstrained}
    \int_\Theta \tilde{\mathrm{L}}(\theta, \xi) \; \mathrm{d}\Pi_{\Theta}(\theta \mid x_{(n)}) = \int_\Theta \mathrm{L}(\mathrm{T}_{\tilde{\Theta}}(\theta), \xi) + \mathrm{dist}(\theta, \mathrm{T}_{\tilde{\Theta}}(\theta)) \; \mathrm{d}\Pi_{\Theta}(\theta \mid x_{(n)}).
  \end{equation}
  For a given decision $\xi$ and samples $\tilde{\theta}^{[i]} \sim \tilde{\Pi}_{\tilde{\Theta}}(\theta \mid x_{(n)})$, the integral in \eqref{eqn:bayes_loss_projected} is numerically calculated as
  \begin{equation} \label{eqn:bayes_risk_proj}
    \tilde{\mathrm{R}}(\xi) = \frac{1}{\mathrm{M}} \sum_{i = 1}^\mathrm{M} \mathrm{L}(\tilde{\theta}^{[i]}, \xi)
  \end{equation}
  and becomes exact as $\mathrm{M} \rightarrow \infty$. Similarly, given samples $\theta^{[i]} \sim \Pi_{\Theta}(\theta \mid x_{(n)})$, the integral in \eqref{eqn:bayes_loss_unconstrained} is numerically calculated as
  \begin{equation} \label{eqn:bayes_risk_unconst}
    \mathrm{R}(\xi) = \frac{1}{\mathrm{M}} \sum_{i = 1}^\mathrm{M} \mathrm{L}(\mathrm{T}_{\tilde{\Theta}}(\theta^{[i]}), \xi) + \mathrm{const}
  \end{equation}
  where the constant is equal to the expectation of $\mathrm{dist}(\theta, \mathrm{T}_{\tilde{\Theta}}(\theta))$ and is invariant in $\xi$. From the definition of $\tilde{\Pi}_{\tilde{\Theta}}(\theta \mid x_{(n)})$, setting $\tilde{\theta}^{[i]} \coloneq \mathrm{T}_{\tilde{\Theta}}(\theta^{[i]})$ provides valid samples from $\tilde{\Pi}_{\tilde{\Theta}}(\theta \mid x_{(n)})$. Hence, \eqref{eqn:bayes_risk_proj} and \eqref{eqn:bayes_risk_unconst} are equal up to a constant, are thus are minimized by the same value $\xi^*$, and the proof of the theorem is obtained.
\end{proof}

\begin{proof} (of \textbf{Corollary~\ref{cor:proj_bayes_risk}})
Since \(\mathrm{T}_{\tilde{\Theta}}\) is non-expansive and
\(\mathrm{T}_{\tilde{\Theta}}(\xi)=\xi\), we have
\[
\|\mathrm{T}_{\tilde{\Theta}}(\theta)-\xi\|
=
\|\mathrm{T}_{\tilde{\Theta}}(\theta)-\mathrm{T}_{\tilde{\Theta}}(\xi)\|
\le
\|\theta-\xi\|.
\]
As \(\ell\) is non-decreasing, it follows that
\[
\mathrm{L}(\mathrm{T}_{\tilde{\Theta}}(\theta),\xi)
=
\ell\!\left(\|\mathrm{T}_{\tilde{\Theta}}(\theta)-\xi\|\right)
\le
\ell\!\left(\|\theta-\xi\|\right)
=
\mathrm{L}(\theta,\xi).
\]
Integrating both sides with respect to \(\Pi_{\Theta}(\cdot\mid x_{(n)})\), yields the pointwise inequality
\[
\int_{\Theta} \mathrm{L}(\mathrm{T}_{\tilde{\Theta}}(\theta),\xi)\,
d\Pi_{\Theta}(\theta\mid x_{(n)})
\le
\int_{\Theta} \mathrm{L}(\theta,\xi)\,
d\Pi_{\Theta}(\theta\mid x_{(n)}).
\]
Minimising both sides over $\xi \in \Xi$, yields the standard Bayes risk on the right-hand side. By Theorem~\ref{the:decision}, the left-hand side is equal to the Bayes risk under the projected posterior.
\end{proof}

\begin{proof} \textbf{(of Theorem~\ref{the:wasserstein})}
  From \cite{villani2009optimal}, a proof to Theorem~\ref{the:wasserstein} is obtained when both $\tilde{\Pi}_{\tilde{\Theta}}(\theta \mid x_{(n)})$ and $ \Pi_\Theta(\theta \mid x_{(n)})$ are valid probability measures on $\Theta$, and by showing that $\mathrm{T}_{\tilde{\Theta}}$ in \eqref{eqn:proj_operator} is a pushforward map between $\Pi_\Theta(\theta \mid x_{(n)})$ and $\nu$, is unique for all $\theta \in \Theta$, and minimizes Monge's formulation of \eqref{eqn:wasserstein},
  \begin{equation} \label{eqn:monge}
  \mathrm{T}_{\tilde{\Theta}} = \underset{\mathrm{T}}{\arg \min} \int_{\Theta} \norm{\theta - \mathrm{T}(\theta)}^2 \; \mathrm{d}\mu(\theta).
  \end{equation}
  The unconstrained and projected posterior distributions are measurable as defined in the main body of the text. Assumption~\ref{ass:map} ensures uniqueness of the projection, and as a consequence $\mathrm{T}_{\tilde{\Theta}}$ is defined as the pushforward map in \eqref{eqn:proj_posterior}. Finally, $\mathrm{T}_{\tilde{\Theta}}$ minimizes \eqref{eqn:monge} as $\norm{\cdot}^2$ is convex and, by definition, $\mathrm{T}_{\tilde{\Theta}}$ minimizes the norm in \eqref{eqn:proj_operator}.
\end{proof}

\begin{proof} \textbf{(of Lemma~\ref{lem:radon_nikodym})}
  Define $\mu$ and $\nu$ as two measures on $\Theta$ such that $\mu$ dominates $\nu$, mathematically expressed as $\nu \ll \mu$. This implies that $\nu$ is absolutely continuous with respect to $\mu$ so that for some $B \in \mathcal{B}$, if $\mu(B) = 0$ then $\nu(B) = 0$. We further define $\tilde{\nu} = \mathrm{T}_{\tilde{\Theta}}(\nu)$ and $\tilde{\mu} = \mathrm{T}_{\tilde{\Theta}}(\mu)$ as the projection of $\nu$ and $\mu$ onto $\tilde{\Theta}$, respectively. Letting $\tilde{B} \in \mathcal{B}_{\tilde{\Theta}}$ such that $\tilde{\mu}(\tilde{B}) = 0$, this implies $\mu\left(\mathrm{T}_{\tilde{\Theta}}^{-1}(\tilde{B})\right) = 0 \Longrightarrow \nu\left(\mathrm{T}_{\tilde{\Theta}}^{-1}(\tilde{B})\right) = 0$ and so $\tilde{\nu}(\tilde{B}) = 0$. Equate $\nu$ and $\tilde{\nu}$ with the unconstrained and projected posterior measures, respectively. By definition, the unconstrained posterior measure is absolutely continuous with respect to a $\sigma$-finite dominating measure $\mu$, and so there exists a $\sigma$-finite dominating measure $\tilde{\mu}$ that dominates $\tilde{\Pi}_{\tilde{\Theta}}(\tilde{B} \mid x_{(n)})$. The remainder of Lemma~\ref{lem:radon_nikodym} is given by the Radon-Nikodym Theorem.
\end{proof}

\begin{proof} \textbf{(of Theorem~\ref{the:projected_prior})}
  As defined in the main text, recall that $\tilde{\pi}_{\tilde{\Theta}}(\theta)$ is the prior density on $\tilde{\Theta}$ that we seek, $\mathrm{p}(x_{(n)} \mid \theta)$ is the likelihood, $\tilde{\pi}_{\tilde{\Theta}}(\theta \mid x_{(n)})$ is the projected posterior density, and $\tilde{\mu}$ is a $\sigma$-finite reference measure on $\tilde{\Theta}$ that dominates the posterior (and hence prior) measure $\tilde{\pi}_{\tilde{\Theta}}(\tilde{B} \mid x_{(n)})$ for $\tilde{B} \in \mathcal{B}_{\tilde{\Theta}}$. From Bayes' rule, we seek some $\tilde{\pi}_{\tilde{\Theta}}(\theta)$ so that 
  \begin{equation} \label{eqn:empirical_posterior}
    \tilde{\pi}_{\tilde{\Theta}}(\theta \mid x_{(n)}) = \frac{\mathrm{p}(x_{(n)} \mid \theta) \tilde{\pi}_{\tilde{\Theta}}(\theta)}{\int_{\tilde{\Theta}} \mathrm{p}(x_{(n)} \mid \theta) \tilde{\pi}_{\tilde{\Theta}}(\theta) \; \mathrm{d}\tilde{\mu}(\theta)}
  \end{equation}
  holds. Consider the prior density
  \begin{equation} \label{eqn:empirical_prior}
    \tilde{\pi}_{\tilde{\Theta}}(\theta) = \frac{\mathrm{p}(x_{(n)} \mid \theta)^{-1} \tilde{\pi}_{\tilde{\Theta}}(\theta \mid x_{(n)})}{\int_{\tilde{\Theta}} \mathrm{p}(x_{(n)} \mid \theta)^{-1} \tilde{\pi}_{\tilde{\Theta}}(\theta \mid x_{(n)}) \; \mathrm{d}\tilde{\mu}(\theta)}
  \end{equation}
  where according to Lemma~\ref{lem:radon_nikodym}, \eqref{eqn:empirical_prior} is well defined, and it is simple to see that it is a probability density as $\int_{\tilde{\theta}} \tilde{\pi}_{\tilde{\Theta}}(\theta) \; \mathrm{d}\tilde{\mu}(\theta) = 1$. Substituting \eqref{eqn:empirical_prior} into the denominator of \eqref{eqn:empirical_posterior} yields
  \begin{align*}
    \int_{\tilde{\Theta}} \mathrm{p}(x_{(n)} \mid \theta) \tilde{\pi}_{\tilde{\Theta}}(\theta) \; \mathrm{d}\tilde{\mu}(\theta) &= \frac{\int_{\tilde{\Theta}} \tilde{\pi}_{\tilde{\Theta}}(\theta \mid x_{(n)}) \; \mathrm{d}\tilde{\mu}(\theta)}{\int_{\tilde{\Theta}} \mathrm{p}(x_{(n)} \mid \theta)^{-1} \tilde{\pi}_{\tilde{\Theta}}(\theta \mid x_{(n)}) \; \mathrm{d}\tilde{\mu}(\theta)} \\
    &= \frac{1}{\int_{\tilde{\Theta}} \mathrm{p}(x_{(n)} \mid \theta)^{-1} \tilde{\pi}_{\tilde{\Theta}}(\theta \mid x_{(n)}) \; \mathrm{d}\tilde{\mu}(\theta)} \\
    &= \frac{\tilde{\pi}_{\tilde{\Theta}}(\theta)}{\mathrm{p}(x_{(n)} \mid \theta)^{-1} \tilde{\pi}_{\tilde{\Theta}}(\theta \mid x_{(n)})},
  \end{align*}
  which substituting back into \eqref{eqn:empirical_posterior} achieves the proof.
\end{proof}

\begin{proof} \textbf{(of Corollary~\ref{cor:emp_bayes})}
  To establish \emph{if and only if} conditions we must show that (1) if $\tilde{\pi}_{\tilde{\Theta}}^{\mathrm{EB}}(\theta; x_{(n)})$ is data-free, then the projected posterior has the standard Bayesian form with prior $q$, and (2) if the projected posterior has the standard Bayesian form with prior $q$, then the empirical Bayes prior in equation~\eqref{eqn:projected_prior} is exactly $q$.

  First, assume $\tilde{\pi}_{\tilde{\Theta}}^{\mathrm{EB}}(\theta; x_{(n)}) = q(\theta)$ for some density $q$ independent of $x_{(n)}$. By definition of $\tilde{\pi}^{\mathrm{EB}}_{\tilde{\Theta}}$ in \eqref{eqn:projected_prior} there exists a normalizing constant $C(x_{(n)})>0$ such that
  \[
  q(\theta)=C(x_{(n)})\,\mathrm{p}(x_{(n)}\mid \theta)^{-1}\,\tilde{\pi}_{\tilde{\Theta}}(\theta\mid x_{(n)}),
  \]
  hence $\tilde{\pi}_{\tilde{\Theta}}(\theta\mid x_{(n)})=\mathrm{p}(x_{(n)}\mid \theta)\,q(\theta)/C(x_{(n)})$.
  Integrating both sides over $\tilde\Theta$ with respect to $\tilde\mu$ and using that $\tilde{\pi}_{\tilde{\Theta}}(\cdot\mid x_{(n)})$ is a probability density yields
  \[
  1=\frac{1}{C(x_{(n)})}\int_{\tilde\Theta}\mathrm{p}(x_{(n)}\mid u)\,q(u)\,\mathrm d\tilde\mu(u),
  \]
  so $C(x_{(n)})=\int_{\tilde\Theta}\mathrm{p}(x_{(n)}\mid u)\,q(u)\,\mathrm d\tilde\mu(u)$ and the displayed Bayes form in Corollary~\ref{cor:emp_bayes} follows, satsfying the first condition.
  
  The second condition is satisfied by noting if $\tilde{\pi}_{\tilde{\Theta}}(\theta\mid x_{(n)}) \propto \mathrm{p}(x_{(n)}\mid \theta)\,q(\theta)$, then substituting into \eqref{eqn:projected_prior} cancels $\mathrm{p}(x_{(n)}\mid \theta)$ and yields $\tilde{\pi}^{\mathrm{EB}}_{\tilde{\Theta}}(\theta; x_{(n)})=q(\theta)$ for $\tilde\mu$-alomst every $\theta$, independent of $x_{(n)}$.
\end{proof}

\begin{proof} \textbf{(of Theorem~\ref{the:consistency})}
  Proposition~6.2 of \cite{ghosal2017fundamentals} states that posterior consistency at $\theta_0$ is achieved if and only if $\Pi_\Theta(\theta \mid x_{(n)}) \leadsto \delta(\theta_0)$ in probability as $n \rightarrow \infty$. This follows from the Portmanteau Theorem. If $\theta_0 \in \tilde{\Theta}$, then similarly $\tilde{\Pi}_{\tilde{\Theta}}(\theta \mid x_{(n)}) \leadsto \delta(\theta_0)$ as the projection $\mathrm{T}_{\tilde{\Theta}}(\theta_0) = \theta_0$.
\end{proof}

\begin{proof} \textbf{(of Theorem~\ref{the:contraction})}
  For any $\theta \in \Theta$ and $\tilde{\theta} \in \tilde{\Theta}$, we have
  \begin{equation}
    \lVert \mathrm{T}_{\tilde{\Theta}}(\theta) - 
    \tilde{\theta} \rVert \leq \norm{\mathrm{T}_{\tilde{\Theta}}(\theta) - 
    \theta} + \lVert \theta - 
    \tilde{\theta} \rVert \leq 2 \lVert \theta - \tilde{\theta} \rVert.
  \end{equation}
  Here, the first inequality is due to the triangle inequality, and the second inequality is due to the definition of the projection operator. Next, due the Assumption of bi-Lipschitz continuity in Assumption~\ref{ass:bi_lipshitz},
  \begin{equation} \label{eqn:bilip_inequality}
    c^{-2} d(\mathrm{T}_{\tilde{\Theta}}(\theta), \tilde{\theta}) \leq c^{-1} \lVert \mathrm{T}_{\tilde{\Theta}}(\theta) - \tilde{\theta} \rVert \leq 2c^{-1} \lVert \theta - \tilde{\theta} \rVert \leq 2d(\theta, \tilde{\theta})
  \end{equation}
  for all $\theta \in \Theta$ and $\tilde{\theta} \in \tilde{\Theta}$. Thus, we may write
  \begin{align}
    \mathrm{T}_{\tilde{\Theta}}^{-1} \{\tilde{\theta} \in \tilde{\Theta} : d(\tilde{\theta}, \theta_0) \geq 2 c^2 M_n \epsilon_n \} &= \left\{\theta \in \Theta : d(\mathrm{T}_{\tilde{\Theta}}(\theta), \theta_0) \geq 2 c^2 M_n \epsilon_n \right\} \label{eqn:proj_set} \\ 
    &\subseteq \left\{\theta \in \Theta : d(\theta, \theta_0) \geq M_n \epsilon_n \right\} \label{eqn:proj_subset}
  \end{align}
  where \eqref{eqn:proj_set} is by definition of the projection operator, and \eqref{eqn:proj_subset} is due to the first and last expressions in \eqref{eqn:bilip_inequality}. Therefore,
  \begin{align*}
    \tilde{\Pi}_{\tilde{\Theta}}\left(\tilde{\theta} \in \tilde{\Theta} : d(\tilde{\theta}, \theta_0) > 2 c^2 M_n \epsilon_n \mid x_{(n)}\right) &= \Pi_\Theta\left(\mathrm{T}_{\tilde{\Theta}}^{-1}\{\tilde{\theta} \in \tilde{\Theta} : d(\tilde{\theta}, \theta_0) > 2 c^2 M_n \epsilon_n \} \mid x_{(n)}\right) \\
    &= \Pi_\Theta\left(\theta \in \Theta : d(\mathrm{T}_{\tilde{\Theta}}(\theta), \theta_0) > 2 c^2 M_n \epsilon_n \mid x_{(n)}\right) \\
    &\leq \Pi_\Theta\left(\theta \in \Theta : d(\theta, \theta_0) > M_n \epsilon_n \mid x_{(n)}\right) \\
    &\rightarrow 0 \text{ in probability for every } M_n \rightarrow \infty.
  \end{align*}
  \end{proof}

  \begin{lemma} \label{lem:bernstein1}
    Given the assumptions in Theorem~\ref{the:bernstein}, and denote by $\tilde{\Theta}^c = \Theta \backslash \tilde{\Theta}^\circ$ the complement of the interior of $\tilde{\Theta}$,
    \begin{equation}
      \mathbb{E}_{\mathrm{P}(x_{(n)} \mid \theta_0)}\left[\mathcal{N}\left(\tilde{\Theta}^c; \hat{\theta}_n, \frac{1}{n} \mathcal{I}^{-1}(\theta_0)\right)\right] \rightarrow 0 \quad \mathrm{as} \quad n \rightarrow \infty,
    \end{equation}
    where, here, $\mathcal{N}\left(\tilde{\Theta}^c; \hat{\theta}_n, \frac{1}{n} \mathcal{I}^{-1}(\theta_0)\right)$ is the measure on $\tilde{\Theta}^c$ parameterized by a Normal distribution with mean $\hat{\theta}_n$ and variance $\frac{1}{n} \mathcal{I}^{-1}(\theta_0)$.
  \end{lemma}
  
  \begin{proof} \textbf{(of Lemma~\ref{lem:bernstein1})}
    All expectations in the following proof are with respect to $\mathrm{P}(x_{(n)} \mid \theta_0)$; this dependency is suppressed in the notation for simplicity. Define a sample $z \sim \mathcal{N}\left(\hat{\theta}_n, \frac{1}{n} \mathcal{I}^{-1}(\theta_0)\right)$, and denote by $\hat{\mathbb{P}}_n(\cdot)=\mathbb{P}(\cdot \mid x_{(n)})$ the conditional, data-dependent law of $z$. Thus
    \begin{equation} \label{lem:lem2_equality}
      \mathbb{E}\left[\mathcal{N}\left(\tilde{\Theta}^c; \hat{\theta}_n, \frac{1}{n} \mathcal{I}^{-1}(\theta_0)\right)\right] = \mathbb{E}\left[\hat{\mathbb{P}}_n(z \in \tilde{\Theta}^c)\right].
    \end{equation}
    Define $\mathbb{B}(\theta_0, \epsilon) = \{\theta \in \Theta : \norm{\theta - \theta_0} < \epsilon\}$ as an open ball around $\theta_0$ for some $\epsilon > 0$ such that $\mathbb{B}(\theta_0, \epsilon) \subset \tilde{\Theta}$. Partition
    \begin{equation} \label{eqn:lemma2_exp}
      \mathbb{E}\left[\hat{\mathbb{P}}_n(z \in \tilde{\Theta}^c)\right] = \mathbb{E}\left[\hat{\mathbb{P}}_n(z \in \tilde{\Theta}^c)\mathds{1}\{\hat{\theta}_n \in \mathbb{B}(\theta_0, \epsilon)\}\right] + \mathbb{E}\left[\hat{\mathbb{P}}_n(z \in \tilde{\Theta}^c)\mathds{1}\{\hat{\theta}_n \notin \mathbb{B}(\theta_0, \epsilon)\}\right].
    \end{equation}
    First consider the first term on the right-hand side of the equality. For every $\hat{\theta}_n \in \mathbb{B}(\theta_0, \epsilon)$ we may define an open ball $\mathbb{B}(\hat{\theta}_n, \delta) \subset \tilde{\Theta}$ for some $\delta > 0$. Due to consistency, as $\mathbb{E}[\hat{\mathbb{P}}_n(z \notin \mathbb{B}(\hat{\theta}_n, \delta))] \rightarrow 0$ for $n \rightarrow \infty$, thus $\mathbb{E}\left[\hat{\mathbb{P}}_n(z \in \tilde{\Theta}^c)\mathds{1}\{\hat{\theta}_n \in \mathbb{B}(\theta_0, \epsilon)\}\right] \rightarrow 0$ for $n \rightarrow \infty$ as $\mathbb{B}(\hat{\theta}_n, \delta) \cap \tilde{\Theta}^c = 0$. Next consider the last term of \eqref{eqn:lemma2_exp},
    \begin{align*}
      \mathbb{E}\left[\hat{\mathbb{P}}_n(z \in \tilde{\Theta}^c)\mathds{1}\{\hat{\theta}_n \notin \mathbb{B}(\theta_0, \epsilon\})\right] &\leq \mathbb{E}\left[\mathds{1}\{\hat{\theta}_n \notin \mathbb{B}(\theta_0, \epsilon)\}\right] \\
      &= \hat{\mathbb{P}}_n(\hat{\theta}_n \notin \mathbb{B}(\theta_0, \epsilon)) \rightarrow 0
    \end{align*}
    for $n \rightarrow \infty$ as $\sqrt{n}(\hat{\theta}_n - \theta_0) \overset{\mathcal{D}}{\rightarrow} \mathcal{N}(0, \mathcal{I}^{-1}(\theta_0))$ and so $\hat{\theta}_n$ converges weakly to a Dirac measure at $\theta_0$. Thus $\mathbb{E}\left[\hat{\mathbb{P}}_n(z \in \tilde{\Theta}^c)\right] \rightarrow 0$ as $n \rightarrow \infty$ and a proof to the Lemma via \eqref{lem:lem2_equality} is obtained.
  \end{proof}

  \begin{lemma} \label{lem:bernstein2}
    Given the assumptions in Theorem~\ref{the:bernstein}, the expectation of the unconstrained posterior measure over $\tilde{\Theta}^c$,
    \begin{equation*}
      \mathbb{E}_{\mathrm{P}(x_{(n)} \mid \theta_0)}\left[\Pi_\Theta(\tilde{\Theta}^c \mid x_{(n)})\right] \rightarrow 0
    \end{equation*}
    as $n \rightarrow \infty$ and where $\tilde{\Theta}^c$ is as defined in Lemma~\ref{lem:bernstein1}.
  \end{lemma}
  
  \begin{proof} \textbf{(of Lemma~\ref{lem:bernstein2})}
    All expectations in the following proof are with respect to $\mathrm{P}(x_{(n)} \mid \theta_0)$; this dependency is suppressed in the notation for simplicity. By definition of the total variation norm
    \begin{equation*}
      \mathbb{E}\left[\left|\Pi_\Theta(\tilde{\Theta}^c \mid x_{(n)}) - \mathcal{N}\left(\tilde{\Theta}^c; \hat{\theta}_n, \frac{1}{n}\mathcal{I}^{-1}(\theta_0)\right)\right|\right] \leq \mathbb{E}\norm{\Pi_\Theta(\theta \mid x_{(n)}) - \mathcal{N}\left(\hat{\theta}_n, \frac{1}{n} \mathcal{I}^{-1}(\theta_0)\right)},
    \end{equation*}
    and so by Assumption~\ref{ass:bernstein},
    \begin{equation*}
      \mathbb{E}\left[\left|\Pi_\Theta(\tilde{\Theta}^c \mid x_{(n)}) - \mathcal{N}\left(\tilde{\Theta}^c; \hat{\theta}_n, \frac{1}{n}\mathcal{I}^{-1}(\theta_0)\right)\right|\right] \rightarrow 0 \quad \text{as} \quad n \rightarrow \infty.
    \end{equation*}
    Lemma~\ref{lem:bernstein1} establishes
    \begin{equation*}
      \mathbb{E}\left[\left|\mathcal{N}\left(\tilde{\Theta}^c; \hat{\theta}_n, \frac{1}{n}\mathcal{I}^{-1}(\theta_0)\right)\right|\right] \rightarrow 0 \quad \text{as} \quad n \rightarrow \infty
    \end{equation*}
    and so $\mathbb{E}\left[\Pi_\Theta(\tilde{\Theta}^c \mid x_{(n)})\right] \rightarrow 0$ as $n \rightarrow \infty$ and a proof to the Lemma is obtained.
  \end{proof}

  \begin{proof} \textbf{(of Theorem~\ref{the:bernstein})}
    In the proof that follows, assume all expectations are with respect to ${\mathrm{P}(x_{(n)} \mid \theta_0)}$ and all norms are the total variation norm. For simplicity, we have suppressed this in the notation. Via the triangle inequality, we bound 
    \begin{equation} \label{eqn:bernstein_inequality}
      \begin{split}
        \mathbb{E}&\norm{\tilde{\Pi}_{\tilde{\Theta}}(\theta \mid x_{(n)}) - \mathcal{N}\left(\theta_0, \frac{1}{n} \mathcal{I}^{-1}(\theta_0)\right)} \leq \\ 
        &\mathbb{E}\norm{\tilde{\Pi}_{\tilde{\Theta}}(\theta \mid x_{(n)}) - \Pi_\Theta(\theta \mid x_{(n)})} + \mathbb{E}\norm{\Pi_\Theta(\theta \mid x_{(n)}) - \mathcal{N}\left(\theta_0, \frac{1}{n} \mathcal{I}^{-1}(\theta_0)\right)}.
      \end{split}
    \end{equation}
    We establish a proof to the theorem by showing that both terms on the right-hand side of the inequality converge to zero as $n \rightarrow \infty$. First, we note that the final term of \eqref{eqn:bernstein_inequality} converges to zero by Assumption~\ref{ass:bernstein}. Next, as $\Pi_\Theta(\theta \mid x_{(n)}) = \tilde{\Pi}_{\tilde{\Theta}}(\theta \mid x_{(n)})$ for $\theta \in \tilde{\Theta}^\circ$
    \begin{equation*}
      \mathbb{E}\norm{\tilde{\Pi}_{\tilde{\Theta}}(\theta \mid x_{(n)}) - \Pi_\Theta(\theta \mid x_{(n)})} \leq \mathbb{E}\left[\Pi_\Theta(\tilde{\Theta}^c \mid x_{(n)})\right].
    \end{equation*}
    From Lemma~\ref{lem:bernstein2}, $\mathbb{E}\left[\Pi_\Theta(\tilde{\Theta}^c \mid x_{(n)})\right] \rightarrow 0$ as $n \rightarrow \infty$ and so
    \begin{equation*}
      \mathbb{E}\norm{\tilde{\Pi}_{\tilde{\Theta}}(\theta \mid x_{(n)}) - \Pi_\Theta(\theta \mid x_{(n)})} \rightarrow 0
    \end{equation*}
    as $n \rightarrow \infty$ and a proof to the theorem is obtained.
  \end{proof}

  \begin{proof} (of \textbf{Theorem~\ref{the:bernstein_manifold}})
  Let $(U,\varphi)$ be a local chart on $\tilde{\Theta}$ with $\varphi:U\subset\mathbb R^k\to\tilde{\Theta}$ and $\varphi(0)=\theta_0$. Write $u=\varphi^{-1}$ on $\varphi(U)$ and define the measurable map
  \[
  g:\Theta\to U,\qquad g(\theta) \coloneq (u\circ T_{\tilde{\Theta}})(\theta).
  \]
  For any probability measure $\mu$ on $(\Theta,\mathcal B(\Theta))$, define its pushforward through $g$ by
  \[
  (g_\sharp \mu)(B) \;:=\; \mu\big(\{\theta\in \Theta: g(\theta)\in B\}\big),\qquad B\in\mathcal B(U).
  \]
  By definition, the projected posterior in chart coordinates is the pushforward of the ambient posterior,
  \[
  \tilde{\Pi}_{\tilde{\Theta}}^u(\cdot\mid x_{(n)}) \;=\; g_\sharp \Pi_\Theta(\cdot\mid x_{(n)}).
  \]
  For any two probability measures $\mu,\nu$ on $\Theta$, total variation is contractive under measurable maps,
  \begin{align*}
  \|g_\sharp\mu-g_\sharp\nu\|_{\mathrm{TV}}
  &=\sup_{B\in\mathcal B(U)} |(g_\sharp\mu)(B)-(g_\sharp\nu)(B)| \\
  &=\sup_{B\in\mathcal B(U)} |\mu(g^{-1}(B))-\nu(g^{-1}(B))| \\
  & \le \sup_{A\in\mathcal B(\Theta)} |\mu(A)-\nu(A)|
  =\|\mu-\nu\|_{\mathrm{TV}}.
  \end{align*}
  The inequality arises from noting that as $g$ is measurable, for each $B \in \mathcal{B}(U)$, $A \coloneq g^{-1}(B) \in \mathcal{B}(\Theta)$. Thus the collection $\{g^{-1}(B) : B \in \mathcal{B}(U)\} \subset \mathcal{B}(\Theta)$. Setting $\mu=\Pi_\Theta(\cdot\mid x_{(n)})$ and $\nu=\mathcal N(\hat\theta_n,n^{-1}\mathcal I(\theta_0)^{-1})$ yields
  \[
  \big\|
  \tilde{\Pi}_{\tilde{\Theta}}^u(\cdot\mid x_{(n)})
  -
  g_\sharp \mathcal N\!\big(\hat\theta_n,n^{-1}\mathcal I(\theta_0)^{-1}\big)
  \big\|_{\mathrm{TV}}
  \le
  \big\|
  \Pi_\Theta(\cdot\mid x_{(n)})
  -
  \mathcal N\!\big(\hat\theta_n,n^{-1}\mathcal I(\theta_0)^{-1}\big)
  \big\|_{\mathrm{TV}}.
  \]
  Taking expectations under $\mathrm P(x_{(n)}\mid\theta_0)$ and invoking Assumption~\ref{ass:bernstein} gives
  \[
  \mathbb E_{\mathrm P(x_{(n)}\mid\theta_0)}
  \big\|
  \tilde{\Pi}_{\tilde{\Theta}}^u(\cdot\mid x_{(n)})
  -
  g_\sharp \mathcal N\!\big(\hat\theta_n,n^{-1}\mathcal I(\theta_0)^{-1}\big)
  \big\|_{\mathrm{TV}}
  \to 0,
  \]
  as $n\to\infty$.
  \end{proof}

\begin{proof} (of \textbf{Proposition~\ref{prop:frob_spiked_projection}})
Fix $i$ and write $\Sigma = \Sigma^{(i)}$ for brevity. Let $\Sigma = U \,\mathrm{diag}(\alpha_1,\dots,\alpha_p)\, U^\top$, $\alpha_1 \ge \cdots \ge \alpha_p,$ be its spectral decomposition. Let $\tilde\Sigma \in \mathcal{M}_r$ with representation
\[
\tilde\Sigma = V \Lambda V^\top + \sigma^2 I_p,
\qquad
V \in \mathrm{St}(p,r), \quad \Lambda = \mathrm{diag}(\lambda_1,\dots,\lambda_r), \quad \lambda_j \ge 0.
\]
Expanding the Frobenius norm,
\[
\|\Sigma -(V\Lambda V^\top+\sigma^2I_p)\|_F^2
=
\|\Sigma \|_F^2
+
\|V\Lambda V^\top+\sigma^2I_p\|_F^2
-
2\,\mathrm{tr}\!\bigl(\Sigma V\Lambda V^\top\bigr)
-
2\sigma^2 \mathrm{tr}(\Sigma ).
\]
For fixed $(\Lambda,\sigma^2)$, the only term depending on $V$ is $-2\,\mathrm{tr}\!\bigl(\Sigma V\Lambda V^\top\bigr)$. Note that $
\|V\Lambda V^\top+\sigma^2I_p\|_F^2
=
\sum_{j=1}^r \lambda_j^2 + 2\sigma^2 \sum_{j=1}^r \lambda_j + p\sigma^4$. Therefore minimising over $V\in\mathrm{St}(p,r)$ is equivalent to maximising $\mathrm{tr}\!\bigl(\Lambda V^\top \Sigma  V\bigr)$.
Define $P := V\Lambda V^\top$, which has rank at most $r$, and non-zero eigenvalues $\lambda_1,\dots,\lambda_r$. By von Neumann's trace inequality,
\[
\mathrm{tr}\!\left(\Sigma P\right)
\le
\sum_{j=1}^r \alpha_j \lambda_j
\]
with equality if and only if $\Sigma$ and $P$ share singular vectors. Thus, $V = U_r$ and by orthogonal invariance of the Frobenius norm,
\begin{align*}
\|\Sigma - \tilde\Sigma\|_F^2
&=
\|U^\top \Sigma U - U^\top \tilde\Sigma U\|_F^2 \\
&=
\left\|
\mathrm{diag}(\alpha_1,\dots,\alpha_p)
-
\mathrm{diag}(\lambda_1+\sigma^2,\dots,\lambda_r+\sigma^2,\sigma^2,\dots,\sigma^2)
\right\|_F^2, \\
&=
\sum_{j=1}^r (\alpha_j - (\lambda_j+\sigma^2))^2
+
\sum_{j=r+1}^p (\alpha_j - \sigma^2)^2.
\end{align*}
Fixing $\sigma^2$, and minimising with respect to $\lambda_j$ and subject to $\lambda_j \ge 0$ yields the solution $\hat{\lambda}_j(\sigma^2) = \max(\alpha_j - \sigma^2,0)$. Substituting back yields a univariate convex problem in $\sigma^2$,
\[
f(\sigma^2)
=
\sum_{j=1}^r \bigl(\alpha_j - \sigma^2 - \hat{\lambda}_j(\sigma^2)\bigr)^2
+
\sum_{j=r+1}^p (\alpha_j - \sigma^2)^2,
\]
where the first term is zero by definition of $\hat{\lambda}_j(\sigma^2)$. Thus, $\hat\sigma^2 =\frac{1}{p-r}\sum_{j=r+1}^p \alpha_j$ and $\hat\lambda_j = \max(\alpha_j - \hat\sigma^2,\,0)$, completing the proof.
\end{proof}

\spacingset{1}
\end{document}